\documentclass[%
preprint,
amsmath,amssymb,
aps,
pre,
showkeys]{revtex4-2}
\usepackage{siunitx}
\usepackage{graphicx}
\usepackage{dcolumn}
\usepackage{bm}
\usepackage{hyperref}
\usepackage{verbatim} 

\begin{document}


\title{Persistent homology of Hamiltonian orbits with applications to beam physics: Euler--Betti classification and topological chaos indicators}
\author{D. Iglesias Tinoco}
\email{diglesiastinoco@niu.edu}
\author{B. Erd\'elyi}%
\email{berdelyi@niu.edu}
\affiliation{
Department of Physics, Northern Illinois University, DeKalb, Illinois, USA}

\date{\today}

\begin{abstract}
Estimating the dynamic aperture of circular particle accelerators typically
requires expensive long-term tracking. Using classical homology, we
construct a two-dimensional theoretical classification of regular and
chaotic orbits and use persistent homology for its numerical
implementation on short orbit segments. In four dimensions, we introduce
topological chaos indicators based on persistent homology and use
normal-form theory to apply the classification to two symplectic-plane
projections. The methods reproduce the expected organization of regular
and chaotic motion in several canonical Hamiltonian systems. Applied to
symplectified one-turn maps of a nonlinear lattice model of the Integrable Optics Test Accelerator,
the results obtained from the first 200 turns identify the central
connected stability region and its surrounding transition to chaos. They are consistent with the short-term
dynamic aperture of the electron configuration obtained from $10^4$
turns and the long-term dynamic aperture of the proton configuration
obtained from $10^8$ turns. For the configurations studied, persistent homology of short orbit segments predicts the dynamic aperture obtained by substantially longer tracking, potentially reducing the need for exhaustive simulations in studies of existing accelerators and the design of future facilities. The methods could also complement classical chaos indicators and modern approaches based on artificial intelligence.
\end{abstract}

\keywords{Dynamic aperture, beam physics, long-term stability, Hamiltonian chaos, topological data analysis, persistent homology}
\maketitle


\section{\label{sec:intro}Introduction}

Understanding and optimizing long-term single-particle stability in the presence of nonlinearities, imperfections, and resonances is a central objective in accelerator beam physics because unstable motion can shorten beam lifetime, limit achievable intensity and brightness, and lead to unwanted radiation or activation of accelerator components.
In many regimes of interest, charged-particle motion through a periodic focusing lattice
admits an accurate Hamiltonian description, often formulated as a symplectic map over one
turn~\cite{Berz1999MapMethods}. Stability questions can therefore be posed within the framework of periodic Hamiltonian dynamics.

A practical and widely used stability proxy is the \emph{dynamic aperture} (DA): the origin-connected region
of initial conditions (ICs) in real space for which particles remain bounded or survive over a prescribed
number of turns or integration steps. Depending on
the machine, particle species, and physics timescale of interest, this may range from
approximately $10^3$ turns for transient phenomena~\cite{Scandale1995DynamicAperture}
to $10^4$ turns for electron storage rings and upwards of $10^9$ turns for long-term proton
tracking.  Despite decades of development, reliable DA prediction and optimization
remain challenging because stability boundaries can be strongly influenced by weak chaos,
resonant transport, and long-lived sticky motion, often necessitating extensive tracking
studies across large parameter spaces~\cite{Scandale1995DynamicAperture,Gareyte1989DA}. For example, in our implementation, directly tracking approximately $10^3$ ICs for $10^8$ turns required nearly six days on a high-performance computing cluster; $10^9$ turns would require roughly two months at the same rate, making a uniform direct scan impractical.

This motivates diagnostics that can extract meaningful stability information from short trajectory segments. Topological data analysis (TDA) provides tools for extracting shape information from finite data sets. Among these tools, persistent homology (PH) is especially useful because it tracks connected components, loops, and voids across a range of geometric scales~\cite{EdelsbrunnerHarer2010CompTopo,CarlssonVJ2021TDA}, with stability properties under suitable perturbations of the input data~\cite{turkes2022on}. In the present setting, a finite early segment of a Hamiltonian orbit is treated as a topological space and PH is used to quantify the topology of the sampled orbit. In contrast to the long-term tracking calculation above, the PH-based diagnostics for the same set of ICs were computed in from 200-turn orbit segments in under one hour on a cloud-computing platform.

Rather than replacing classical chaos indicators, such topological diagnostics can complement them by providing early-time information that can be used alongside longer-trajectory methods such as Lyapunov indicators, frequency map analysis (FMA), the Smaller Alignment Index (SALI), the Reversibility Error Method (REM), or the Hurst exponent~\cite{ChaosIndicators,GiovannozziPhysRevE.107.064209,SALI,Borin}. In addition, recent artificial-intelligence-based approaches, including convolutional neural networks and physics-informed neural networks, have been used for chaos classification~\cite{Chaos-ML1,Chaos-DeepLearning}. The resulting
topology-based descriptors may also be incorporated into hybrid artificial-intelligence
frameworks, for example by vectorizing persistence information or storing it as image-like
input for neural-network classifiers~\cite{Adams2017PersistenceImages}.

In this work, we introduce topology-based orbit classifiers and chaos indicators aimed at
beam-physics applications. We begin with idealized classifications based on Betti numbers and the Euler--Poincar\'e characteristic~\cite{poincare1895analysis,siersma2012poincare, EdelsbrunnerHarer2010CompTopo, Hatcher}, and construct the associated numerical implementation via PH. The resulting PH information is summarized through Euler-characteristic-type quantities; the necessary PH background is reviewed in Appendix~\ref{sec:PH_appendix}. Related
PH-based orbit classification has been developed for magnetic field-line Poincar\'e maps in
perturbed tokamak models, where orbit types are assigned using thresholding rules on PH
summaries and the resulting performance can depend on the chosen parameters and the
symplectic map under study~\cite{BohlsenRobinsHole2025FieldLinePH}. 

The paper is structured as follows. In Sec.~\ref{sec:Classification}, we develop a 2D Euler--Betti orbit classifier and introduce both intrinsic and extrinsic 4D PH diagnostics. Because Betti numbers describe the connected-component, loop, and void structure of an orbit rather than the specific analytic form of a map, the 2D classifier is naturally adaptable across symplectic maps and does not require map-specific threshold tuning. Its numerical implementation still depends on finite-data choices such as the sampling window, metric, and choice of complex. The intrinsic indicators are based on Euler-characteristic-type summaries of PH lifespans. Their scales are local to the map and sampled region and therefore require calibration against regular and chaotic reference orbits. The extrinsic construction forms two symplectic-plane projections of each 4D orbit in uncoupled or normal-form coordinates and applies the 2D classifier separately to them.

In Sec.~\ref{sec:Validation}, we test the 2D classifier on several canonical Hamiltonian systems using short orbit segments from 2D maps and 2D Poincar\'e sections of a 4D system. It distinguishes invariant tori, islands, and chaotic layers and is benchmarked against classical chaos indicators. We also test the intrinsic indicators on a 4D symplectic map. In Sec.~\ref{sec:IOTA}, we combine these diagnostics into a PH workflow for estimating DA and apply it to nonlinear IOTA lattice configurations for electrons and protons. With its reconfigurable lattice, capability to operate with multiple particle species and role as a platform for testing concepts relevant to future accelerator systems~\cite{Antipov2017IOTA}, IOTA is well suited to this beam-physics application of the PH framework. The resulting estimates are benchmarked against substantially longer tracking.
Finally, Sec.~\ref{sec:Conclusion and Outlook} summarizes the main findings, limitations, and possible extensions, including integration with machine-learning methods.

\section{\label{sec:Classification}Topological Classification of Hamiltonian Orbits}

\subsection{\label{sec:2DClass} 2D classification}

\subsubsection{\label{sec:Theory}Idealized classification}

We begin with an idealized topological classification of orbit types observed on
two-dimensional Poincar\'e sections. The purpose of this classification is not to provide
a unique definition of Hamiltonian chaos, but rather to identify homological templates
that guide the construction of a practical finite-data classifier. 

\begin{figure*}[htpb]
  \centering
\includegraphics[width=\textwidth]{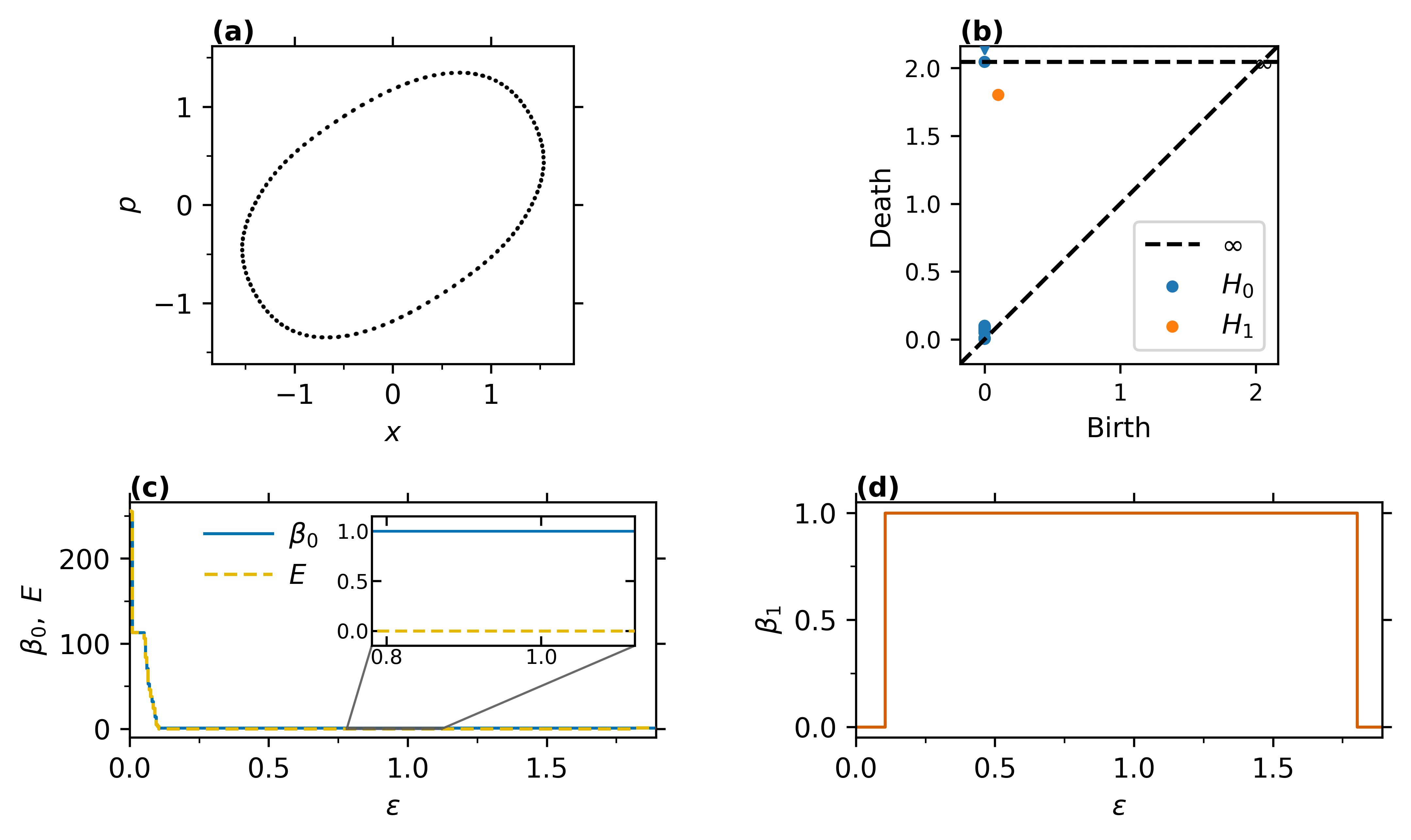}
  \caption{Representative of a torus-class orbit and its PH diagnostics. (a) Finite sampled orbit in a Poincar\'e section, approximating an idealized closed loop, and (b) its persistence diagram. (c) $\beta_0(\epsilon)$, $E(\epsilon)=\beta_0(\epsilon)-\beta_1(\epsilon)$, and (d) $\beta_1(\epsilon)$. The single contiguous run with $\beta_1=1$ and $E_{\mathrm{mode}}=0$ gives the numerical torus signature.
}
  \label{fig:Torus}
\end{figure*}

\begin{figure*}[htpb]
  \centering
\includegraphics[width=\textwidth]{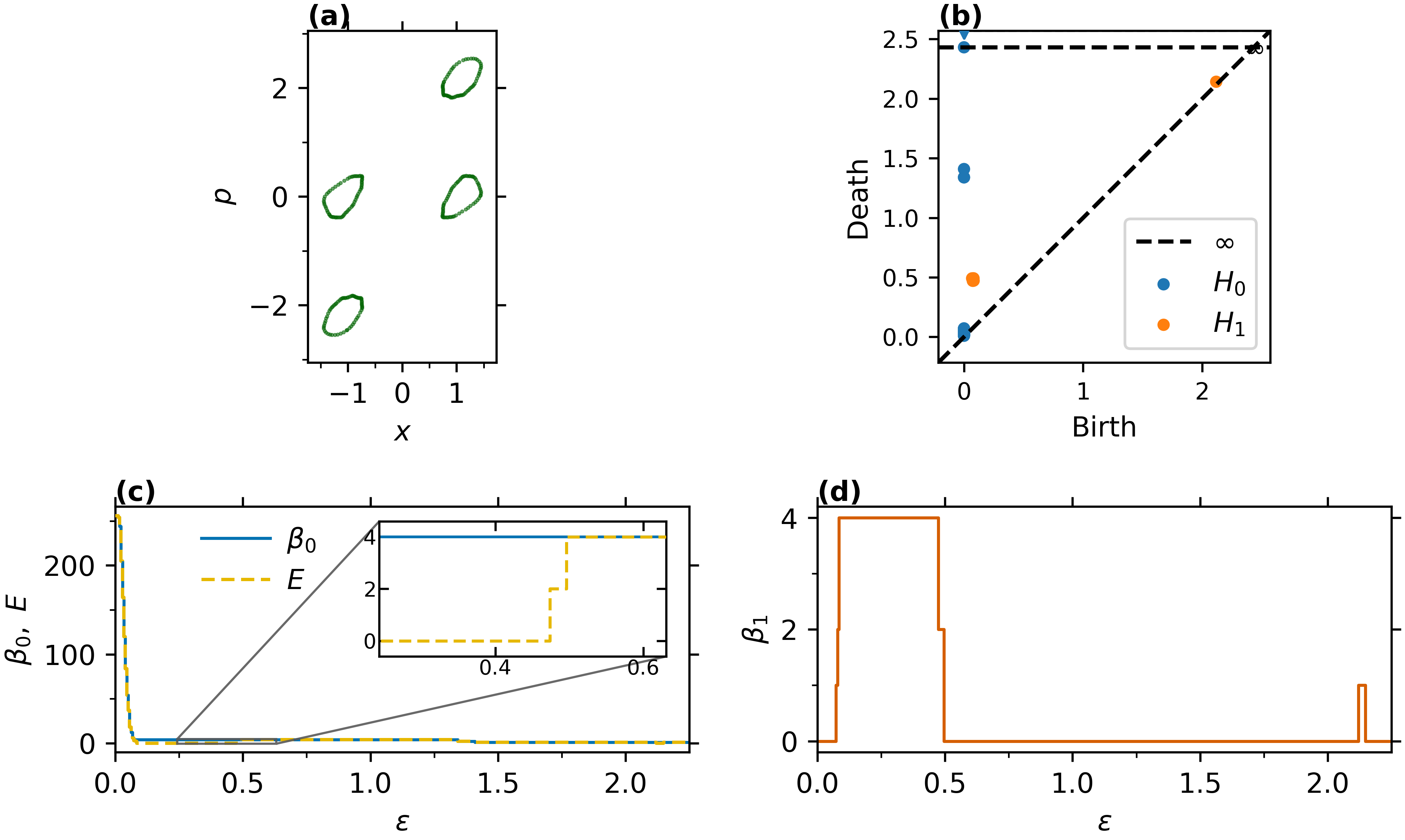}
  \caption{Representative of an island-class orbit of order $k=4$ and its PH diagnostics. (a) Finite sampled orbit in a Poincar\'e section and (b) its persistence diagram. (c) $\beta_0(\epsilon)$, $E(\epsilon)=\beta_0(\epsilon)-\beta_1(\epsilon)$, and (d) $\beta_1(\epsilon)$. The separated peaks in $\beta_1(\epsilon)$ represent the individual island loops at finer scales and a filtration-induced loop after the four loops connect at a larger scale. Here, $\beta_{1,\max}=E_{\mathrm{mode}}=\beta_{0,\mathrm{mode}}=4$, giving the numerical island signature.}
  \label{fig:Island}
\end{figure*}

\begin{figure*}[ht]
  \centering
\includegraphics[width=\textwidth]{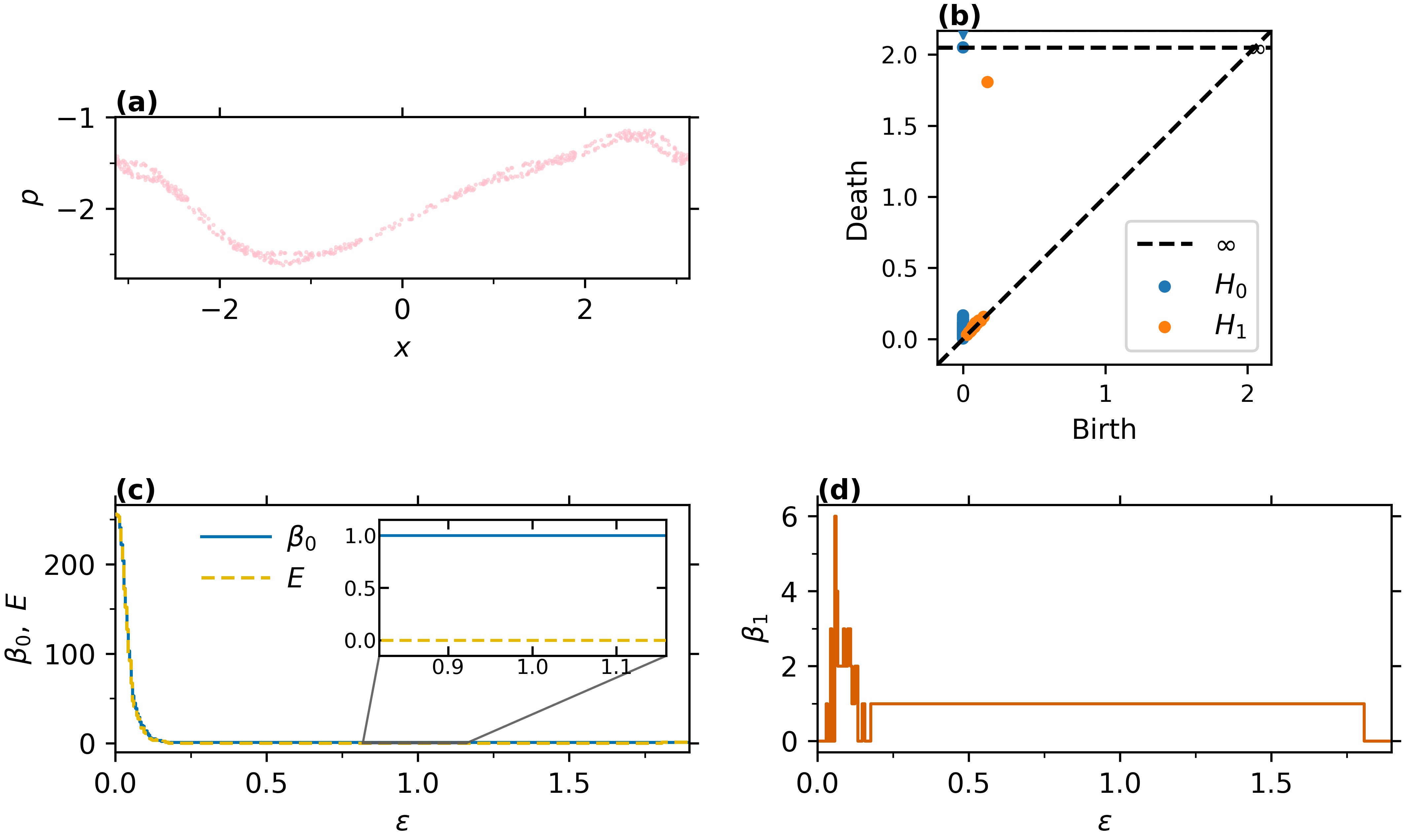}
  \caption{Representative of a thin-chaos T1 orbit and its PH diagnostics. (a) Finite sampled orbit in a periodic Poincar\'e section and (b) its persistence diagram; the left and right boundaries panel (a) are identified. (c) $\beta_0(\epsilon)$, $E(\epsilon)=\beta_0(\epsilon)-\beta_1(\epsilon)$, and (d) $\beta_1(\epsilon)$. Additional short-lived $H_1$ classes near the diagonal of the persistence diagram produce $\beta_{1,\max}>1$, while $E_{\mathrm{mode}}=0$, giving the numerical thin-chaos T1 signature.}
  \label{fig:ThinChaosT1}
\end{figure*}

\begin{figure*}[htpb]
  \centering
\includegraphics[width=\textwidth]{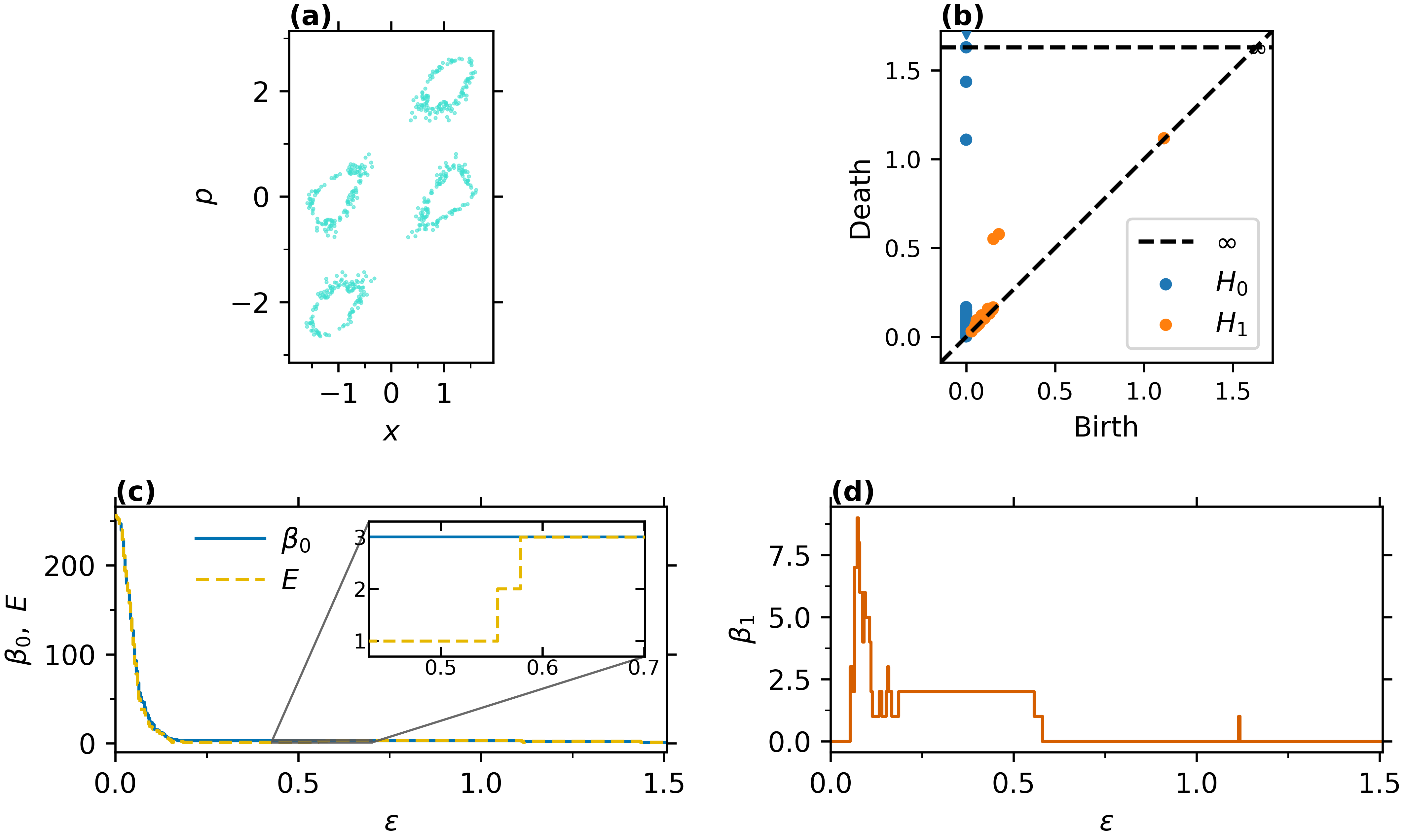}
  \caption{Representative a thin-chaos T2 perturbation of an order-$4$ island chain and its PH diagnostics. (a) Finite sampled orbit in a Poincar\'e section  and (b) its persistence diagram . (c) $\beta_0(\epsilon)$, $E(\epsilon)=\beta_0(\epsilon)-\beta_1(\epsilon)$, and (d) $\beta_1(\epsilon)$. Additional short-lived $H_1$ classes near the diagonal of the persistence diagram give $\beta_{1,\max}=9$, whereas $E_{\mathrm{mode}}=\beta_{0,\mathrm{mode}}=3$, so the numerical island equality is not satisfied and the orbit is classified as thin chaos T2.}
  \label{fig:ThinChaosT2}
\end{figure*}

\begin{figure*}[htpb]
  \centering
\includegraphics[width=\textwidth]{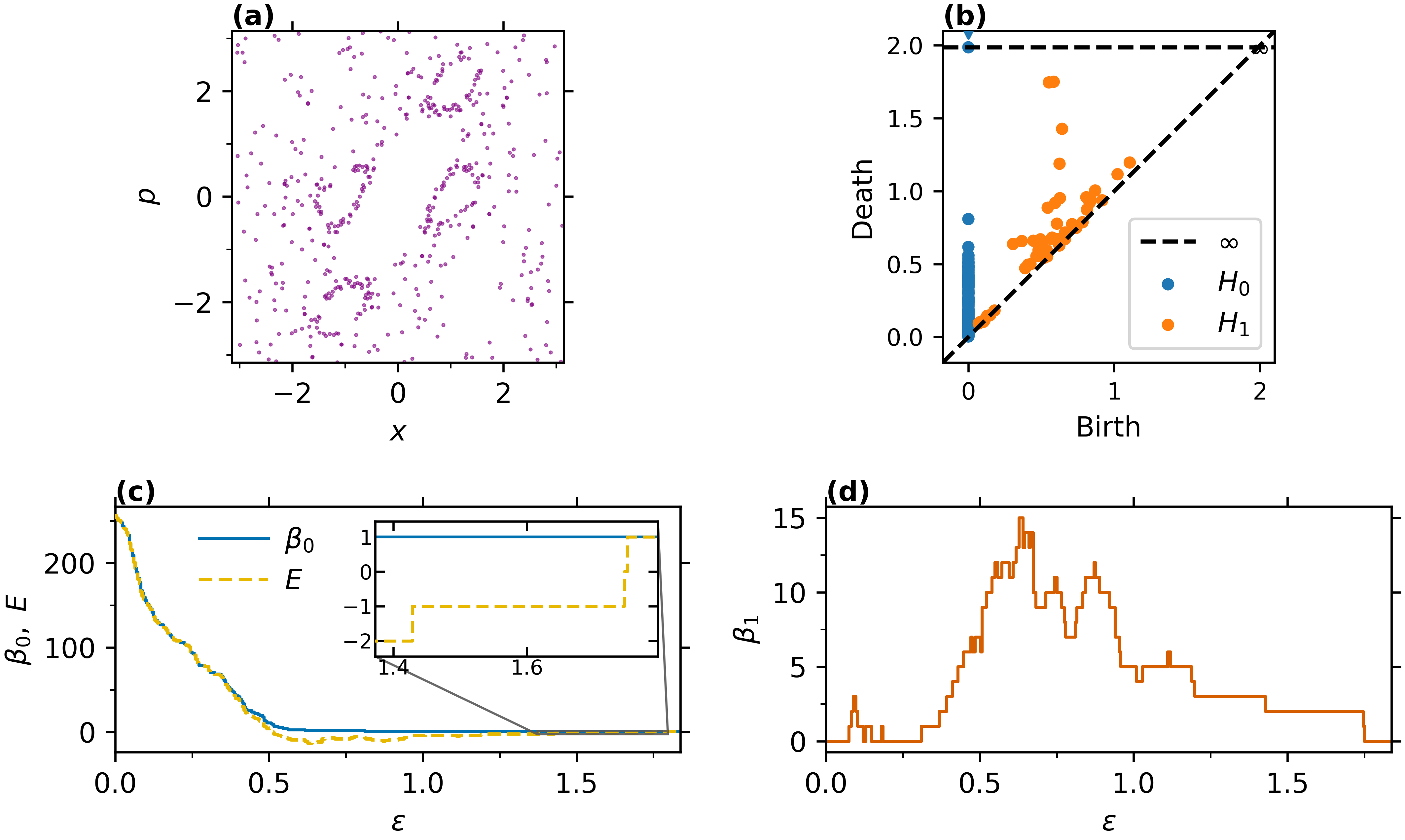}
  \caption{Representative of a thick-chaos orbit and its PH diagnostics. (a) Finite sampled orbit in a Poincar\'e section and (b) its persistence diagram. (c) $\beta_0(\epsilon)$, $E(\epsilon)=\beta_0(\epsilon)-\beta_1(\epsilon)$, and (d) $\beta_1(\epsilon)$. The numerous $H_1$ classes in the persistence diagram produce an extended range with $\beta_1(\epsilon)>\beta_0(\epsilon)$ and $E_{\mathrm{mode}}=-1<0$, giving the numerical thick-chaos signature.}
  \label{fig:ThickChaos}
\end{figure*}

Throughout this subsection, $\beta_n$ denotes the $n$th classical Betti number of an idealized orbit, reviewed in Appendix~\ref{sec:Betti,Euler-Poincare,Kunneth}. The (a) panels of Figs.~\ref{fig:Torus}--\ref{fig:ThickChaos} show finite orbit samples drawn from the examples considered later. Here, they serve only to illustrate the geometry of the corresponding idealized orbit topologies; the particular map and IC do not enter the definition of the classes.  The remaining (b)-(d) panels display the associated persistence diagram (PD) and the numerical curves $\beta_0(\epsilon)$, $\beta_1(\epsilon)$, and $E(\epsilon)$, where $\epsilon$ is the filtration scale. These PH diagnostics are defined in Appendix~\ref{sec:PH_persistent} and analyzed in the following subsection.

The first nontrivial class is that of quasi-periodic motion on an invariant torus.
On a two-dimensional Poincar\'e section, a representative orbit appears as a closed loop,
with one connected component and one independent loop, i.e.,\ $(\beta_0,\beta_1)=(1,1)$,
as geometrically illustrated in Fig.~\ref{fig:Torus} (a).

Next consider an island chain of order $k>1$, for which an orbit traces $k$ disjoint closed loops on a 2D Poincar\'e section, one around each island in the chain. Topologically, these loops may be idealized as $k$ disjoint copies of the torus representative, so that the corresponding Betti numbers are
$(\beta_0,\beta_1)=k(1,1)=(k,k)$.
 Thus the orbit has $k$ connected components and
$k$ independent loops.  Figure~\ref{fig:Island} (a) shows an order 4 island chain obtained numerically. We do not separately classify islands of order
$k=1$, since such orbits do not arise in the beam-physics applications that motivate the
present work, although they do appear in plasma-physics settings
\cite{BohlsenRobinsHole2025FieldLinePH}. In that context, order-$1$ islands can be
distinguished from tori using an enclosure function that determines whether the loop
encloses the magnetic axis.

We next introduce two idealized classes of thin chaos, denoted T1 and T2. Thin chaos of
type T1 is defined as a perturbation of $\beta_1$ of the torus class,
\begin{equation}
(\beta_0,\beta_1)=(1,1)\mapsto(1,p), \qquad p>1,
\end{equation}
so that the orbit remains connected while developing additional loop structure. In the
ideal picture, this corresponds to a loop-like set with multiple holes, similar to a
necklace.  Figure.~\ref{fig:ThinChaosT1} (a) illustrates this orbit topology; since the plot is periodic, the left and right boundaries are
identified.

Similarly, thin chaos of type T2 is defined as a perturbation of $\beta_1$ of an island chain of order
$k$,
\begin{equation}
(\beta_0,\beta_1)=(k,k)\mapsto(k,km), \qquad m>1,
\end{equation}
so that the number of connected components is preserved while the number of loops within
each component increases. Equivalently, T2 may be viewed as $k$ disjoint copies of a
T1-type structure: $k(1,m)$. A representative of this class is shown in 
Fig.~\ref{fig:ThinChaosT2} (a).

Finally, we define thick chaos as a stronger departure from these thin-chaos templates.
From the T1 point of view, thick chaos may be regarded as a transition
\begin{equation}
(1,p)\mapsto(1,q), \qquad q>p,
\end{equation}
in which the number of loops increases substantially. From the T2 point of view, it may
equivalently be regarded as a transition
\begin{equation}
(k,km)\mapsto(1,q), \qquad q>km,
\end{equation}
in which disconnected components merge into a single connected chaotic layer while the
number of loops increases.
Figure~\ref{fig:ThickChaos} (a) contains only a limited number of iterates, so the idealized thick chaotic layer is only partially resolved.  

\begin{table}[htpb]
\caption{Idealized two-dimensional Euler--Betti orbit classification.}
\label{tab:theory_EB}
\begin{ruledtabular}
\begin{tabular}{lcccc}
Topological Space 
$X$
& $\beta_0$
& $\beta_1$
& $(\beta_0,\beta_1)$
& $\chi(X)=\beta_0-\beta_1$ \\
\hline
Fixed point of order $m$ & $m$ & $0$   & $(m,0)$ & $m>0$ \\
Torus (loop)                  & $1$ & $1$   & $(1,1)$ & $0$ \\
Island of order $k$ ($k>1$)   & $k$ & $k$   & $(k,k)$ & $0$ \\
Thin chaos T1                 & $1$ & $p>1$ & $(1,p)$ & $1-p<0$ \\
Thin chaos T2                 & $k$ & $km$  & $(k,km)$ & $k(1-m)<0$ \\
Thick chaos                   & $1$ & $q\gg 1$ & $(1,q)$ & $1-q\ll 0$ \\
\end{tabular}
\end{ruledtabular}
\end{table}

The idealized classifications are summarized in
Table~\ref{tab:theory_EB} along with their corresponding Euler--Poincar\'e characteristic $\chi$; see Eq.~\eqref{Euler-Poincare}. Fixed points are included in the idealized classification for
completeness. For finite computed orbit data, however, finite precision
or an IC that only approximates a theoretical fixed point
can produce a small nonconstant orbit cloud. For this reason, the
2D numerical classification introduced in the next subsection does not
include a separate fixed-point class. Known trivial fixed points may
instead be handled directly from the orbit coordinates.
The last column provides an idealized topological chaos indicator:
periodic motion has positive $\chi$, quasi-periodic motion has $\chi=0$,
thin chaotic layers have negative $\chi$, and thick chaos has the most
negative values. 

\subsubsection{\label{sec:Numerics}Numerical classification}

We now translate the idealized topological classification, expressed in terms of classical Betti numbers, into a practical numerical classifier using summaries of the persistent Betti curves of finite orbit samples. In particular, we use the modal value
(most frequent value) of the Euler--Betti curve $E(\epsilon)$
[see Eq.~\eqref{Euler-Betti}], denoted $E_{\mathrm{mode}}$, together
with simple features of the piecewise Betti step functions, such as
$\beta_{1,\max}$. The modes of $E(\epsilon)$ and $\beta_0(\epsilon)$ are evaluated over the sampled filtration
values; when a mode is not unique, the smallest value is selected. These
quantities capture the dominant topology of the orbit point cloud under
finite sampling.

For a torus, $\beta_1(\epsilon)=1$ occurs in a single contiguous run,
and the most frequent value of $E(\epsilon)$ is $0$; see
Fig.~\ref{fig:Torus} (c)-(d). If this topology is weakly perturbed over a small
range of $\epsilon$, several additional short-lived loops may appear
near the diagonal of the PD, as in 
Fig.~\ref{fig:ThinChaosT1} (b). For the examples considered here,
$E_{\mathrm{mode}}$ remains $0$. We then classify the orbit as a torus when
$\beta_{1,\max}=1$ and $\beta_1(\epsilon)=1$ occurs in a single
contiguous run. The remaining cases with $E_{\mathrm{mode}}=0$ are
classified as thin chaos of type T1.

For islands, the idealized signature of $\beta_1(\epsilon)$ is a
double-step structure with a wide separation between peaks; see Fig.~\ref{fig:Island} (d). The first peak represents the individual loops of the island chain.
As $\epsilon$ increases, these loops fill while the $k=4$ components
remain disconnected. At a larger filtration scale, the components
connect around the chain and produce a larger loop, giving the second
peak. This larger loop is induced by the filtration and is not part of
the ideal topology of the original orbit. Numerically, an island is
identified when $
\beta_{1,\max}
=
E_{\mathrm{mode}}
=
\beta_{0,\mathrm{mode}}
>1$.
If $E_{\mathrm{mode}}\geq1$ and $\beta_{1,\max}>0$, but the island
condition is not satisfied, the orbit is classified as thin chaos of
type T2. The main difference between the $H_1$ persistence data for an
island of order $4$ [Fig.~\ref{fig:Island} (b)] and
its corresponding thin-chaos T2 perturbation [Fig.~\ref{fig:ThinChaosT2} (b)] is the presence of many additional classes
near the diagonal, as expected from the idealized classification.

Finally, thick chaos is the easiest case to identify numerically. In the idealized picture, it has the most negative Euler--Poincar\'e characteristic $\chi(X)$. In practice, this corresponds to an Euler--Betti curve $ E(\epsilon)$ for which negative values dominate; see Fig.~\ref{fig:ThickChaos} (c). We therefore classify an orbit as thick chaos whenever $E_{\mathrm{mode}}<0$.

A summary of the numerical classification is given in
Table~\ref{tab:numerical_EB}. We also record $E_{\min}$, defined as the minimum sampled value of $E(\epsilon)$. This quantity is shown in some
figures for exploratory comparison with the idealized classification.

\begin{table}[htpb]
\caption{Numerical two-dimensional Euler--Betti classification. The criteria are applied from top to bottom.}
\label{tab:numerical_EB}
\begin{ruledtabular}
\begin{tabular}{lcc}
Orbit type 
& $\beta_{1,\max}(\epsilon)$ & $E_{\text{mode}}$ \\
\hline
Torus         & $1$, one contiguous run & $0$ \\
Thin chaos T1 & Otherwise & $0$ \\
Island        & $\beta_{0,\text{mode}}>1$ & $\beta_{0,\text{mode}}>1$ \\
Thin chaos T2 & $>0$ & $\geq1$ \\
Thick chaos   & $\text{--}$ & $<0$ \\
Unclassified  & $\text{--}$ & Otherwise \\
\end{tabular}
\end{ruledtabular}
\end{table}

\subsection{\label{sec:4DClass} 4D classification}

\subsubsection{\label{sec:4DIdealized}Idealized product topologies}

For the 2D maps considered in Secs.~\ref{sec:Chirikov} and \ref{sec:Harper}, the numerical classifier introduced above is applied directly to individual orbits using a few hundred iterates. For the 4D H\'enon--Heiles flow in Sec.~\ref{sec:Henon}, by contrast, $7.5\times10^4$ integration steps are used to obtain a comparable number of points on a 2D Poincar\'e section. More generally, constructing a sufficiently sampled lower-dimensional section of a higher-dimensional system may therefore require substantially longer trajectories, which the beam-physics applications considered here motivate avoiding. We consequently extend the idealized theory to representative 4D product topologies. The limitations of converting these ideal signatures into a practical classifier motivate the lifespan-based intrinsic indicators introduced in Sec.~\ref{sec:Intrinsic}. Nevertheless, when uncoupled or approximately decoupled coordinates are available, the 2D classifier can still be applied separately to planar projections of 4D orbit segments, providing the complementary extrinsic classification of Sec.~\ref{sec:Extrinsic Classification}.

Product topologies provide an idealized reference framework for
4D orbit structure. The purpose of this construction is not
to provide a complete classification of 4D Hamiltonian
dynamics, but to determine how representative 2D orbit
classes combine when the motion contains two independent degrees of
freedom.

For nonresonant quasi-periodic motion with two independent frequencies, a
single orbit densely covers an invariant two-torus, $S^1 \times S^1 = T^2$.
In the uncoupled case, the two circle factors can be identified naturally
with invariant loops in the two symplectic planes. However, loop-like projections in both symplectic planes do not by
themselves imply that the full 4D orbit densely covers a
two-torus. When the frequencies satisfy a resonance relation, the phases
are no longer independent and the orbit may instead lie on a
1D closed curve embedded in the 4D phase
space. Such resonances have been observed experimentally in the
CERN Super Proton Synchrotron~\cite{Schmidt2024}. 

Since a loop has Betti numbers $(\beta_0,\beta_1)=(1,1)$, the nonresonant product $S^1\times S^1$ has Betti numbers $(\beta_0,\beta_1,\beta_2)=(1,2,1)$  and therefore $\chi(T^2)=1-2+1=0$ just as in the 2D one-torus $T^1=S^1$. The Betti numbers of product spaces follow from the K\"unneth formula in Eq.~\eqref{Kunneth Product}. In the present setting, suppose that the topological spaces
$X$ and $Y$ have 
Betti data $(\beta_0^X,\beta_1^X)$ and
$(\beta_0^Y,\beta_1^Y)$, respectively. Their product then has
\begin{align}
\beta_0(X\times Y)
    &= \beta_0^X\beta_0^Y, \\
\beta_1(X\times Y)
    &= \beta_1^X\beta_0^Y+\beta_0^X\beta_1^Y, \\
\beta_2(X\times Y)
    &= \beta_1^X\beta_1^Y.
\end{align}
Applying these relations to representative 2D orbit classes
from Table~\ref{tab:theory_EB} gives the idealized product examples listed in Table~\ref{tab:theory_EB_4D}, where the fixed-point, island and thick chaos classes are denoted by FP, I and TC respectively.
These examples are not intended to form a complete taxonomy of
4D orbit structure, but rather to provide reference
topological signatures. 

\begin{table}[htpb]
\caption{Examples of idealized four-dimensional classes.}
\label{tab:theory_EB_4D}
\begin{ruledtabular}
\begin{tabular}{lcccc}
Topological space: $X\times Y$
& $\beta_0$
& $\beta_1$
& $\beta_2$
& $\chi$ \\
\hline
FP $\times$ FP: $(m,0)\times(n,0)$
& $mn$ & $0$ & $0$ & $mn>0$ \\
FP $\times~T^1$: $(m,0)\times(1,1)$
& $m$ & $m$ & $0$ & $0$ \\
Two-torus: $(1,1)\times(1,1)$
& $1$ & $2$ & $1$ & $0$ \\
I $\times$ I: $k(1,1)\times m(1,1)$
& $km$ & $2km$ & $km$ & $0$ \\
T1 $\times~T^1$ : $(1,p)\times(1,1)$
& $1$ & $1+p$ & $p$ & $0$ \\
T1 $\times$ T2: $(1,p)\times(k,km)$
& $k$ & $k(m+p)$ & $kpm$  & $>0$ \\
\end{tabular}
\end{ruledtabular}
\end{table}

Table~\ref{tab:theory_EB_4D} shows that the Euler--Poincar\'e characteristic alone loses
information contained in the individual Betti numbers. This follows in
part from the product relation $
\chi(X\times Y)=\chi(X)\chi(Y)$ in Eq.~\eqref{Euler-Poincare Product}. Since a one-torus has $\chi(T^1)=0$, any product containing a $T^1$ factor
also has zero Euler--Poincar\'e characteristic. Thus,
$\mathrm{FP}\times T^1$, $T^2$, and the chaotic product
$\mathrm{T1}\times T^1$ all have $\chi=0$, although their Betti tuples
are different. By contrast, the product of two chaotic classes with
negative Euler--Poincar\'e characteristics may have a positive value. For
$\mathrm{T1}\times\mathrm{T2}$, $\chi(\mathrm{T1})\chi(\mathrm{T2})=k(1-p)(1-m)>0$. Consequently, neither the value nor
the sign of $\chi$ alone provides an unambiguous 4D chaos
indicator.

Nevertheless, the product construction remains useful because the familiar
regular orbit templates considered here exhibit symmetric patterns in their
Betti numbers. A one-torus has the symmetric Betti tuple $(1,1)$ in its
intrinsic dimension, while the nonresonant two-torus and the
island-product class have the symmetric tuples $(1,2,1)$ and
$km(1,2,1)$, respectively. A resonant closed curve embedded in
4D phase space has Betti tuple $(1,1,0)$ when dimensions
zero through two are listed, but intrinsically it remains a
1D loop with tuple $(1,1)$. By contrast, the chaotic product
templates generally depart from these symmetric patterns. For example,
$\mathrm{T1}\times T^1$ has Betti tuple $(1,1+p,p)$, while
$\mathrm{T1}\times\mathrm{T2}$ has tuple $k(1,p+m,pm)$.

For a finite orbit point cloud, however, PH does not
necessarily recover one of these ideal Betti tuples at a single filtration
scale. Even for a sampled two-torus, the two independent loops associated
with $\beta_1=2$ may be born at different values of the filtration
parameter $\epsilon$, while the class associated with $\beta_2=1$ may
appear at a later scale. Consequently, the persistent Betti tuple $
\left(\beta_0(\epsilon),\beta_1(\epsilon),\beta_2(\epsilon)\right) $
need not equal $(1,2,1)$ over a common or sufficiently persistent range of
$\epsilon$. This motivates the lifespan-based Euler-type summaries
introduced in the next subsection and defined formally in
Appendix~\ref{sec:PH_persistent}.

The product viewpoint also remains useful in the extrinsic construction
introduced in Sec.~\ref{sec:Extrinsic Classification}. When the motion is uncoupled, or
can be represented in approximately decoupled normal-form coordinates defined in Sec.~\ref{sec:Extrinsic Classification}, the
2D classifier of Table~\ref{tab:numerical_EB} can be applied
separately to the two symplectic planes and the resulting labels can be
combined. This provides a practical interpretation of the orbit in terms
of its horizontal and vertical behavior, although the projected classes do not
uniquely determine the topology of the full 4D orbit
because projection discards correlations between the two planes.
For the flat-beam case considered in Sec.~\ref{sec:IOTA}, midplane
symmetry~\cite{BelaSymmetries} preserves an invariant plane (in coordinates introduced later,
$y=b=0$). Orbits launched in this plane may therefore be interpreted as
embedded product classes of the form
$\mathcal{C}\times\mathrm{FP}$, where $\mathcal{C}$ may be a fixed point,
torus, island, T1, T2, or TC class in the horizontal plane.

\subsubsection{\label{sec:Intrinsic} Intrinsic topological chaos indicators}

As discussed in the previous subsection, PH need
not recover the complete ideal Betti tuple at a single filtration scale.
We therefore summarize the persistence of homological classes over the
complete filtration. To preserve the alternating structure of the
Euler--Poincar\'e characteristic, we replace the Betti numbers at a fixed
filtration value by lifespan sums computed separately in homological
dimensions zero, one, and two.

For the barcode set $B_n$ in homological dimension $n$, let
$\tau^p(B_n)$ denote the $p$th lifespan sum defined in
Eq.~\eqref{eq:tau_np}. The corresponding intrinsic Euler-type quantity through
homological dimension two is
\begin{equation}
\chi_{\tau^p}
=
\tau^p(B_0)-\tau^p(B_1)+\tau^p(B_2).
\end{equation}
The two quantities considered in this work are $\chi_\tau$, corresponding
to $p=1$, and $\chi_{\tau^2}$, corresponding to $p=2$.
Although this construction is motivated by the theoretical Betti patterns discussed above, the lifespan triple does not necessarily reproduce their
symmetry or relative ordering. In particular, the symmetry of the two-torus Betti tuple does not imply that
$
\left(
\tau^p(B_0),
\tau^p(B_1),
\tau^p(B_2)
\right)
$
must be symmetric, nor does the value $\beta_1=2$ imply that $\tau^p(B_1)$ must be the largest lifespan sum. Betti numbers count the independent homology classes present at a specified filtration scale, whereas $\tau^p(B_n)$ sums the lifespans of all finite persistent classes in dimension $n$ over the complete filtration.

This distinction is particularly important in dimension zero. An orbit
point cloud containing $N$ distinct points initially has $N$ connected
components and therefore $N$ classes in $H_0$. As the filtration parameter
$\epsilon$ increases, these components merge until only one essential
component remains. After omitting the corresponding infinite-persistence
class, the zero-dimensional persistence data $B_0$ contain $N-1$ finite
birth--death pairs. Consequently, $\tau^p(B_0)$ can be the largest entry
of the lifespan triple even for a regular quasi-periodic orbit.

Within a fixed numerical protocol, compact and coherently sampled orbit
clouds generally become connected at smaller filtration scales, whereas
orbit clouds spread over a broader region generally require larger scales
to merge. Thus, $\tau^p(B_0)$ contains information about the connectivity
and spatial dispersion of the sampled orbit. The alternating quantity
$\chi_{\tau^p}$ combines this zero-dimensional contribution with the
lifespan contributions $\tau^p(B_1)$ and $\tau^p(B_2)$ associated with
loops and two-dimensional homology classes, respectively. This
interpretation requires the number of sampled points, coordinate
representation, metric, and sampled phase-space region to be held fixed.

For the orbit clouds considered here, chaotic motion generally produces
greater phase-space spreading. This can increase the lifespans of the
$H_0$ classes recorded in $B_0$, because more widely separated
components merge at larger filtration scales. In some cases, it may also
increase the lifespans of the $H_2$ classes recorded in $B_2$. Since
both terms enter $\chi_{\tau^p}$ with positive signs, they may dominate
the negative contribution $\tau^p(B_1)$ and produce a larger indicator
value. The numerical scale and ordering nevertheless depend on the map,
sampled region, and persistent-homology protocol.

The distinction between the two indicators lies in how the persistence
lifespans are weighted. In $\chi_{\tau}$, each finite lifespan contributes
linearly. In $\chi_{\tau^2}$, each lifespan is squared before the three
contributions are combined. Consequently, persistent (long-lived) classes, represented
by points farther from the diagonal of the PD, receive
greater relative weight, while short-lived classes near the diagonal have
less influence.

For a fixed map, sampled phase-space region, and numerical protocol, we
expect compact regular or quasi-periodic orbit clouds to produce smaller
values of $\chi_{\tau}$ and $\chi_{\tau^2}$. Weakly chaotic orbit clouds
are expected to produce intermediate values, while stronger phase-space
spreading can increase the positive $B_0$ and $B_2$ contributions and
produce larger values. This expected local ordering is summarized in
Table~\ref{tab:intrinsic-ordering}.

\begin{table}[t]
\caption{\label{tab:intrinsic-ordering}
Expected relative behavior of the intrinsic Euler-type persistence summaries. The ordering is local to a fixed map, sampled phase-space region, and persistent-homology protocol, and is tested numerically in Sec.~\ref{sec:CoupledChirikov}.}
\begin{ruledtabular}
\begin{tabular}{lcc}
Orbit type & $\chi_{\tau}$ & $\chi_{\tau^2}$ \\
\hline
Regular or quasi-periodic motion & Lower & Lower \\
Weak or thin chaos & Intermediate & Intermediate \\
Strong or thick chaos & Higher & Higher \\
\end{tabular}
\end{ruledtabular}
\end{table}

The ordering in Table~\ref{tab:intrinsic-ordering} is local rather than universal and
therefore requires calibration within a fixed map and sampled region.
Nevertheless, the indicators are stable with respect to perturbations of
the PDs. For two orbit point clouds $P$ and $Q$, the
difference $\left|
\chi_{\tau^p}(P)-\chi_{\tau^p}(Q)
\right|$
is bounded in terms of the $p$-Wasserstein distances between their
corresponding persistence data $B_n(P)$ and $B_n(Q)$, as shown in
Appendix~\ref{sec:PH_persistent}. Thus, orbit clouds with nearby PDs
have nearby indicator values, even though universal numerical thresholds
for regular and chaotic motion are not expected.

\subsubsection{\label{sec:Extrinsic Classification} Extrinsic numerical classification}

The intrinsic indicators treat each orbit as a point cloud in the full
four-dimensional phase space. As a complementary extrinsic construction, we assume that the orbit can be expressed in uncoupled or
approximately decoupled normal-form coordinates. 

In the uncoupled case, the two
canonical pairs evolve independently; for example, the coupled Chirikov
map introduced later in Eqs.~\eqref{eq:coupled_chirikov_first}--\eqref{eq:coupled_chirikov_last} reduces to two
uncoupled 2D maps when $\xi=0$. The beam-physics maps
considered in Sec.~\ref{sec:IOTA} are generally coupled in the original coordinates, so their orbits are first transformed to normal-form
coordinates, explicit procedure given in Sec.~\ref{sec:Normalization}. A normal-form transformation provides a nonlinear change of
variables that simplifies the underlying symplectic map and, up to a
chosen order, represents the motion as approximately circular (a rotation) in each transformed symplectic plane with
amplitude-dependent frequencies. The normal-form algorithm is described in~\cite{Berz1999MapMethods}.

At iteration $n$, we denote the orbit point in these coordinates by $\bar z_n=(\bar x_n,\bar a_n,\bar y_n,\bar b_n)$, where $(\bar x,\bar a)$ and $(\bar y,\bar b)$ are the first and second
canonical pairs, respectively. We then form the projections
\begin{equation}
\bar z_n^{(X)}=(\bar x_n,\bar a_n),
\qquad
\bar z_n^{(Y)}=(\bar y_n,\bar b_n).
\label{eq}
\end{equation}

The numerical two-dimensional classifier of
Table~\ref{tab:numerical_EB}, supplemented by a preliminary fixed-point
check, is applied independently to each projected data set. The
resulting labels describe the orbit behavior observed in the horizontal
and vertical symplectic planes.

For each projected symplectic plane, the resulting orbit classes are
grouped into four numerical plane labels under the uncoupled or
approximately decoupled assumptions stated above. The classifier first
checks whether the total coordinate spread of the projected orbit is at
most $10^{-12}$; if so, the orbit is identified directly as a fixed
point, without PH. A nonconstant projection satisfying
$E_{\mathrm{mode}}>0$ and $\beta_{1,\max}=0$ is also treated as a
fixed-point case for the purpose of assigning the plane label. Fixed
points, tori, and islands are assigned the label $S$; thin chaos of
types T1 and T2 are assigned the labels $W1$ and $W2$, respectively;
and thick chaos is assigned the label $C$. Here, $S$ is a finite-sample stability label: the observed projected
orbit remains bounded and exhibits a regular topological signature over
the sampled iterations. It does not imply linear or Lyapunov stability
with respect to nearby initial conditions. This correspondence is
summarized in Table~\ref{tab: stability classifier}.

\begin{table}[htpb]
\caption{Short-term stability classifier}
\label{tab: stability classifier}
\begin{ruledtabular}
\begin{tabular}{lc}
Two-dimensional orbit class & Plane label \\
\hline
Fixed point & $S$ \\
Torus & $S$ \\
Island & $S$ \\
Thin chaos T1 & $W1$ \\
Thin chaos T2 & $W2$ \\
Thick chaos & $C$ \\
\end{tabular}
\end{ruledtabular}
\end{table}

The final extrinsic classification is the ordered combination $L_X\times L_Y$,
where $L_X,L_Y$ are the labels obtained from the $(\bar x,\bar a),(\bar y,\bar b)$ projections, respectively.  The ordering
is retained because the horizontal and vertical behaviors need not be the
same; for example, $\mathrm{S}\times\mathrm{W1}$ and
$\mathrm{W1}\times\mathrm{S}$ describe different projected dynamics. Here
the symbol $\times$ denotes the combination of the two plane-level labels
and does not by itself imply that the full 4D orbit has an
exact product topology. The possible combinations are listed in
Table~\ref{tab: extrinsic classifier}.

\begin{table}[htpb]
\caption{Extrinsic four-dimensional labels obtained by combining the
short-term stability labels assigned to the horizontal and vertical
symplectic-plane projections. Rows correspond to the $(\bar x,\bar a)$ plane and
columns to the $(\bar y,\bar b)$ plane.}
\label{tab: extrinsic classifier}
\begin{ruledtabular}
\begin{tabular}{lcccc}
$(\bar x,\bar a)\backslash(\bar y,\bar b)$ & $S$ & $W1$ & $W2$ & $C$ \\
\hline
$S$ & $S\times S$ & $S\times W1$ & $S\times W2$ & $S\times C$ \\
$W1$ & $W1\times S$ & $W1\times W1$ & $W1\times W2$ & $W1\times C$ \\
$W2$ & $W2\times S$ & $W2\times W1$ & $W2\times W2$ & $W2\times C$ \\
$C$ & $C\times S$ & $C\times W1$ & $C\times W2$ & $C\times C$ \\
\end{tabular}
\end{ruledtabular}
\end{table}

\section{\label{sec:Validation}Validation on canonical systems}

To test the Euler--Betti classifier from
Table~\ref{tab:numerical_EB}, we consider the Chirikov standard map
(Sec.~\ref{sec:Chirikov}) and the kicked Harper map
(Sec.~\ref{sec:Harper}) in two dimensions, and the H\'enon--Heiles
Hamiltonian, whose 4D phase space is analyzed through a
2D Poincar\'e section (Sec.~\ref{sec:Henon}). The intrinsic
4D chaos indicators from Table~\ref{tab:intrinsic-ordering} are tested on a coupled Chirikov standard
map (Sec.~\ref{sec:CoupledChirikov}). These systems probe the diagnostics
across a range of phase-space structures relevant to Hamiltonian
dynamics.

The Chirikov standard map is a canonical model of the transition from integrable to
non-integrable behavior and provides a natural first test case because its phase-space
organization is well understood and extensively documented in the literature~\cite{Chirikov1979,Mirella2019,Borin, Zaslavsky2007HamiltonianChaos}. It is also
a standard building block for higher-dimensional symplectic models, for example through
coupled-map constructions~\cite{Lange2014GlobalTori4D}. The kicked Harper map is included both as an additional
benchmark and because of its physical relevance in several applications. Finally, the
H\'enon--Heiles system~\cite{HenonHeiles1964ThirdIntegral} is a classical model for nonlinear Hamiltonian dynamics and is
particularly relevant here because of its recent use in beam-physics-related studies~\cite{HenonHeilesIOTA}.

The coordinate representation used for PH depends on the
topology of the phase space. For maps with periodic coordinates, each
coordinate $u$ of period $L$ is represented by its circle embedding
$
\Phi_L(u)=
\left(
\cos(2\pi u/L),
\sin(2\pi u/L)
\right)$,
and the Euclidean metric is applied after embedding. Thus, the Chirikov
standard map and the kicked Harper map are embedded coordinate-wise using
$L=2\pi$ and $L=1$, respectively, while the coupled Chirikov map is treated
analogously in four dimensions. For the Hénon--Heiles Poincar\'e section,
whose coordinates are nonperiodic, PH is computed directly
using the Euclidean metric on either $(y,p_y)$ or its three-dimensional
embedding $(p_x,y,p_y)$.

\subsection{\label{sec:Chirikov}Chirikov standard map}

We begin with the Chirikov standard map~\cite{Chirikov1979}, using the minus-sign convention of
Ref.~\cite{Borin}:
\begin{align}
p_{n+1} &= p_n - K \sin(x_n) \pmod{2\pi},\\
x_{n+1} &= x_n + p_{n+1} \pmod{2\pi}.
\end{align}
Here $K$ is the nonlinearity parameter controlling the transition from
predominantly regular to increasingly chaotic dynamics. In the
charged-particle interpretation discussed in
Ref.~\cite{Zaslavsky2007HamiltonianChaos}, $K$ is proportional to the
magnitude of the external electric field,
$K\propto|\vec{E}|$.

\begin{figure*}[htpb]
  \centering
\includegraphics[width=\textwidth]{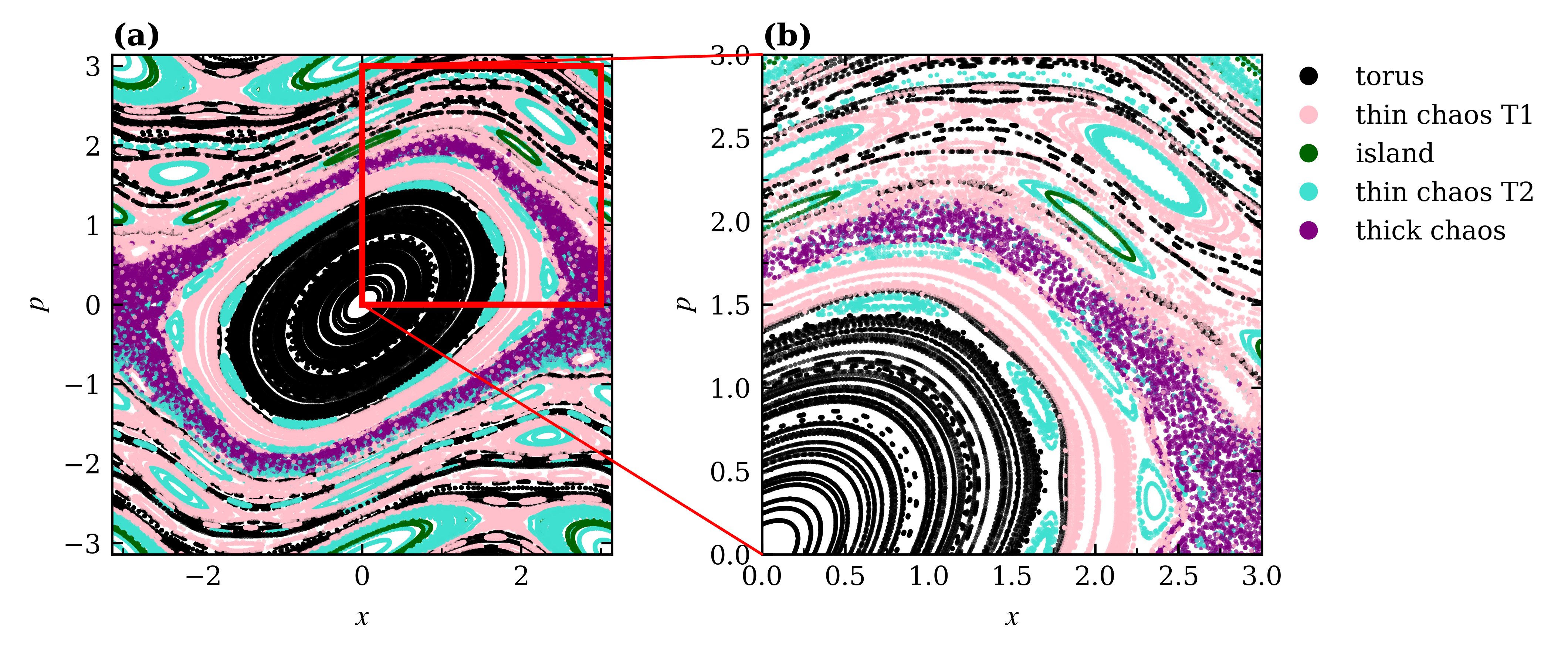}
  \caption{Orbit classification for the Chirikov standard map with $K=0.9$, $2^8$ orbits, $2^9$ points per orbit and $2^8$ points used for classification. (a) Sampled phase-space portrait, orbits are colored according to the Euler--Betti class indicated in the legend. (b) Zoom of the region outlined in red.}
\label{fig:ChirikovAClassColor}
\end{figure*}

\begin{figure*}[htpb]
  \centering
\includegraphics[width=\textwidth]{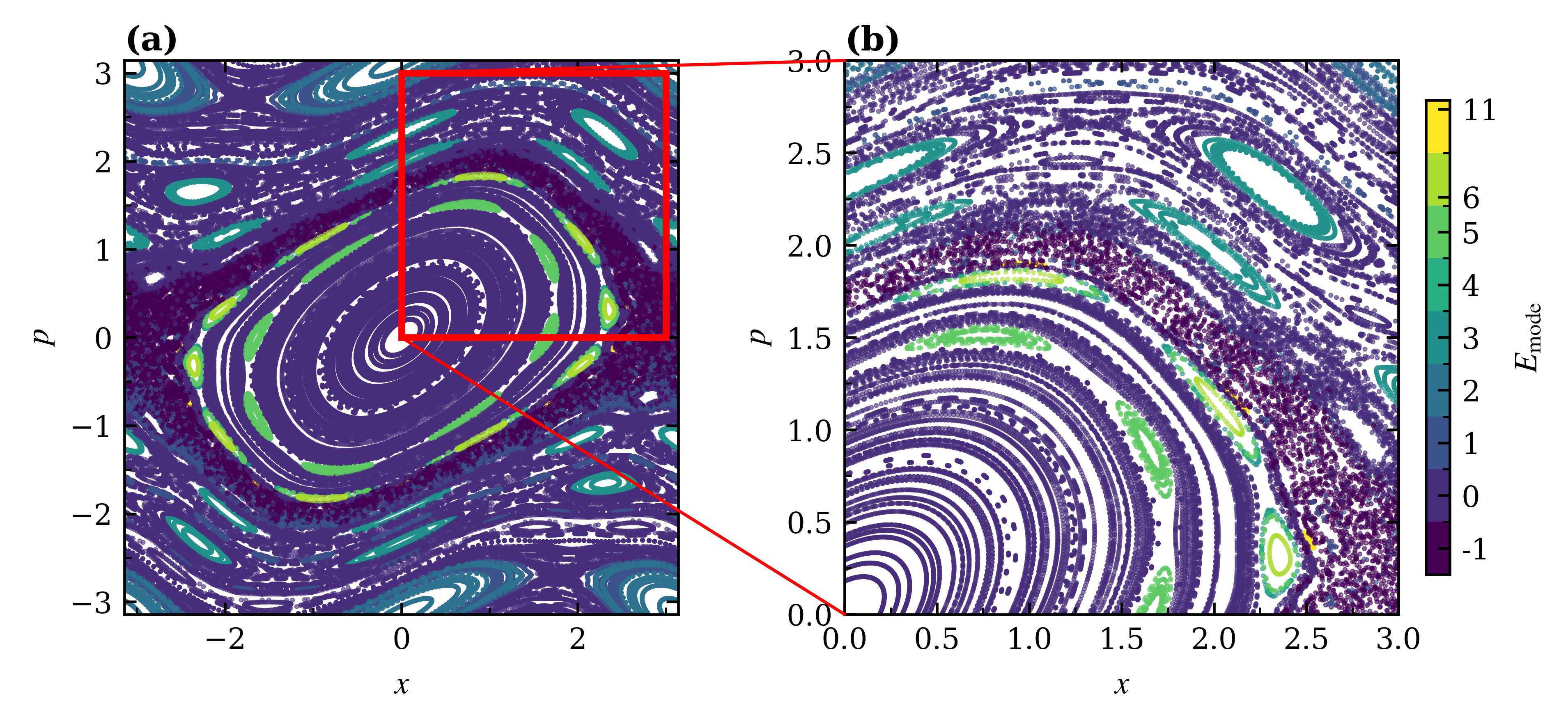}
  \caption{(a) Chirikov standard map phase-space portrait at \(K=0.9\), colored by \(E_{\mathrm{mode}}\), the modal sampled value of \(E(\epsilon)\). (b) Zoom of the region outlined in red.
}
  \label{fig:ChirikovBEColor}
\end{figure*}

\begin{figure*}[htpb]
  \centering
\includegraphics[width=\textwidth]{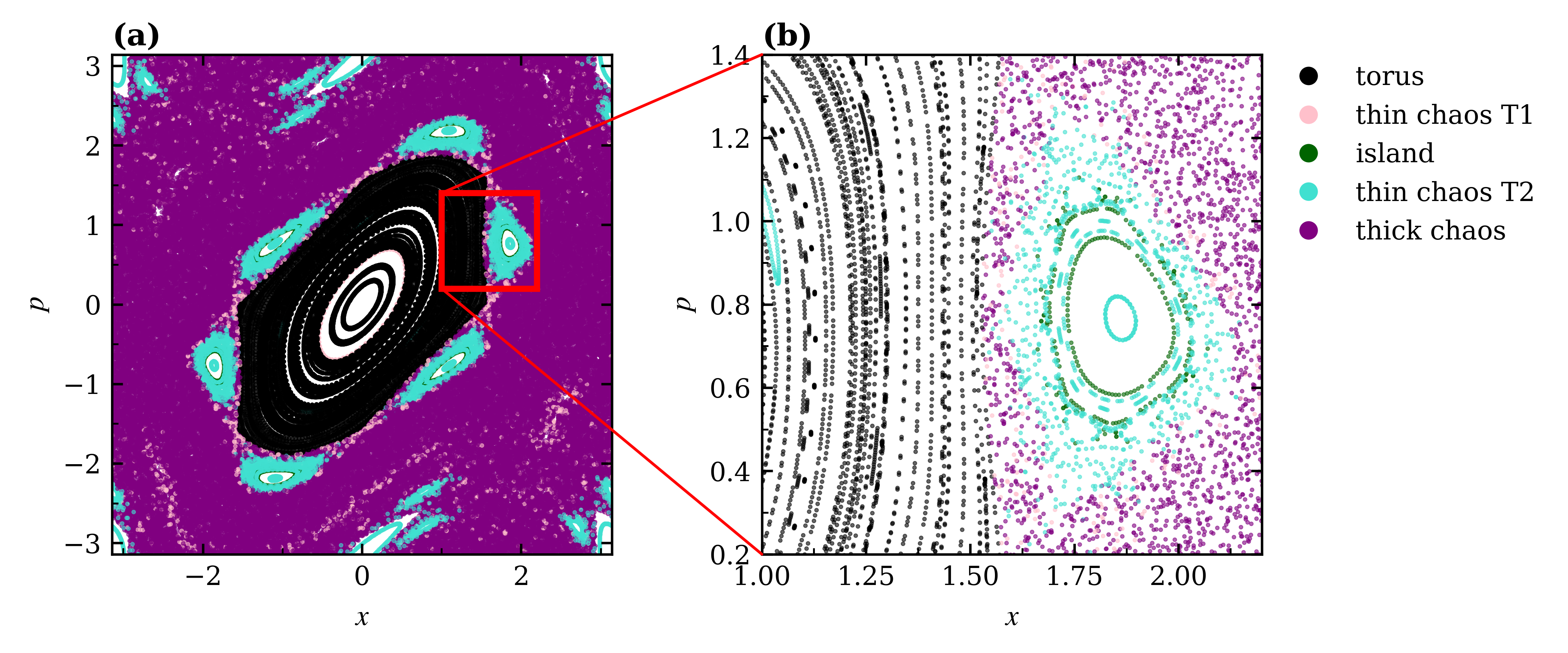}
  \caption{Orbit classification for the Chirikov standard map with $K=1.6$, $2^8$ orbits, $2^9$ points per orbit and $2^8$ points used for classification. (a) Sampled phase-space portrait, orbits are colored according to the Euler--Betti class indicated in the legend. (b) Zoom of the region outlined in red.}
\label{fig:Chirikov1.6ClassColor}
\end{figure*}

\begin{figure*}[htpb]
  \centering
\includegraphics[width=\textwidth]{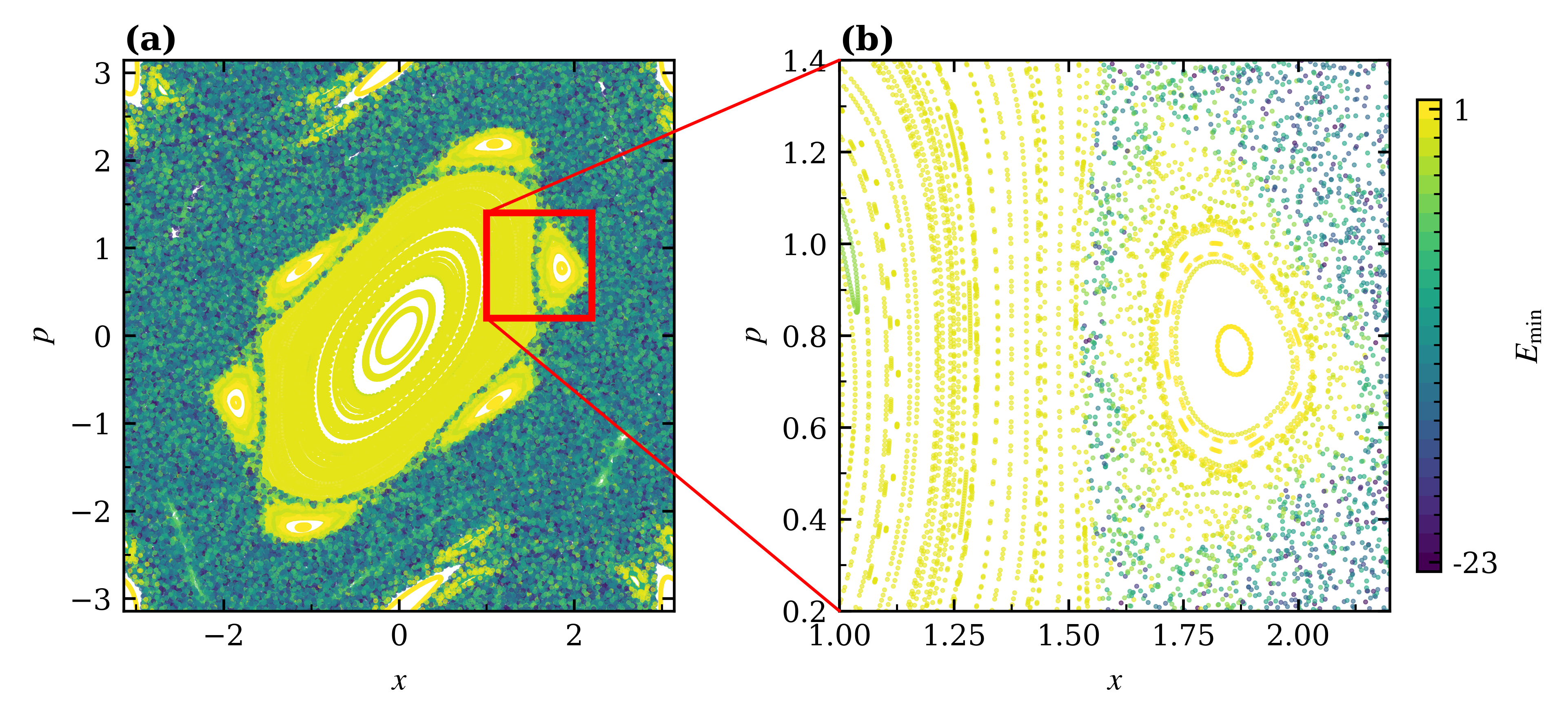}
  \caption{(a) Chirikov standard map phase-space portrait at \(K=1.6\), colored by \(E_{\min}\), the minimum sampled value of \(E(\epsilon)\). (b) Zoom of the region outlined in red.
}
  \label{fig:Chirikov1.6EColor}
\end{figure*}

\begin{figure*}[htpb]
  \centering
\includegraphics[width=\textwidth]{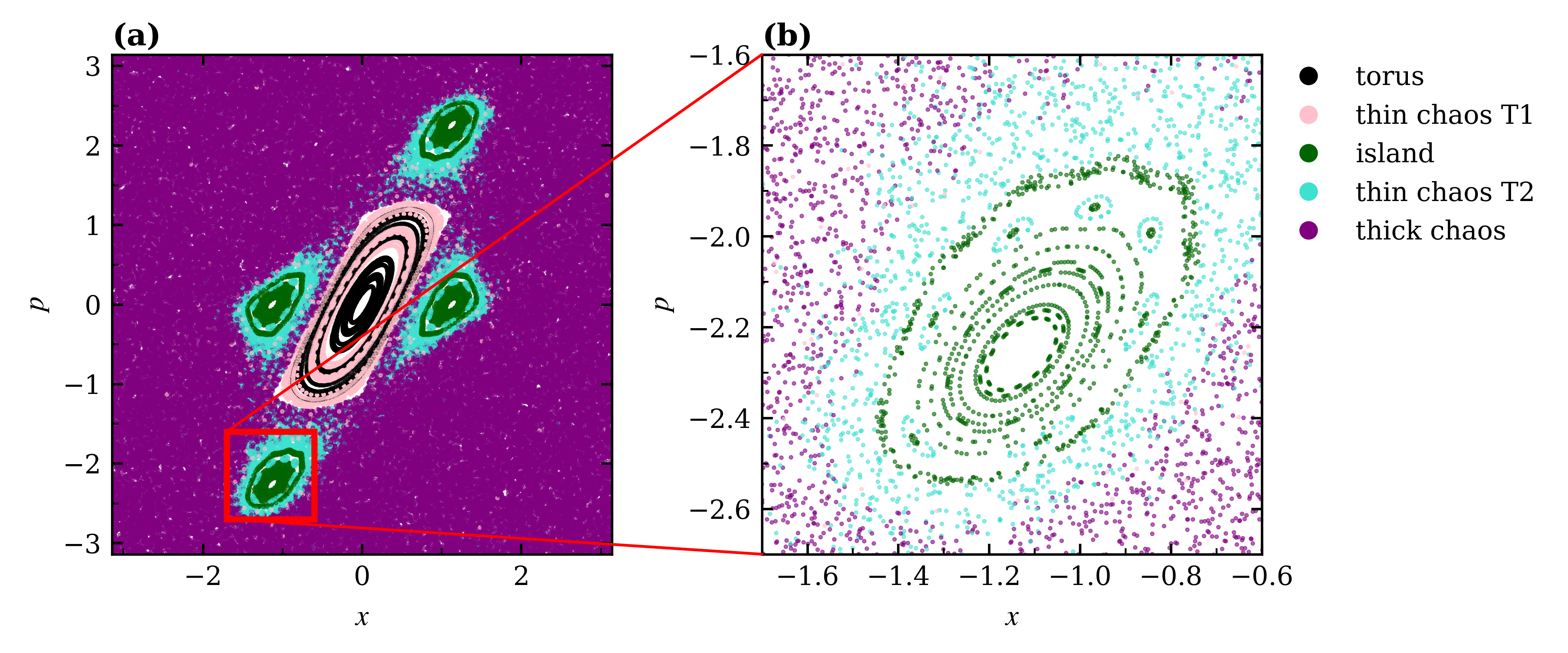}
  \caption{Orbit classification for the Chirikov standard map with $K=2.5$, $400$ orbits, $2^8$ points per orbit and all used for classification. (a) Sampled phase-space portrait, orbits are colored according to the Euler--Betti class indicated in the legend. (b) Zoom of the region outlined in red.}
  \label{fig:Chirikov2.5ClassColor}
\end{figure*}

\begin{figure*}[htpb]
  \centering
\includegraphics[width=\textwidth]{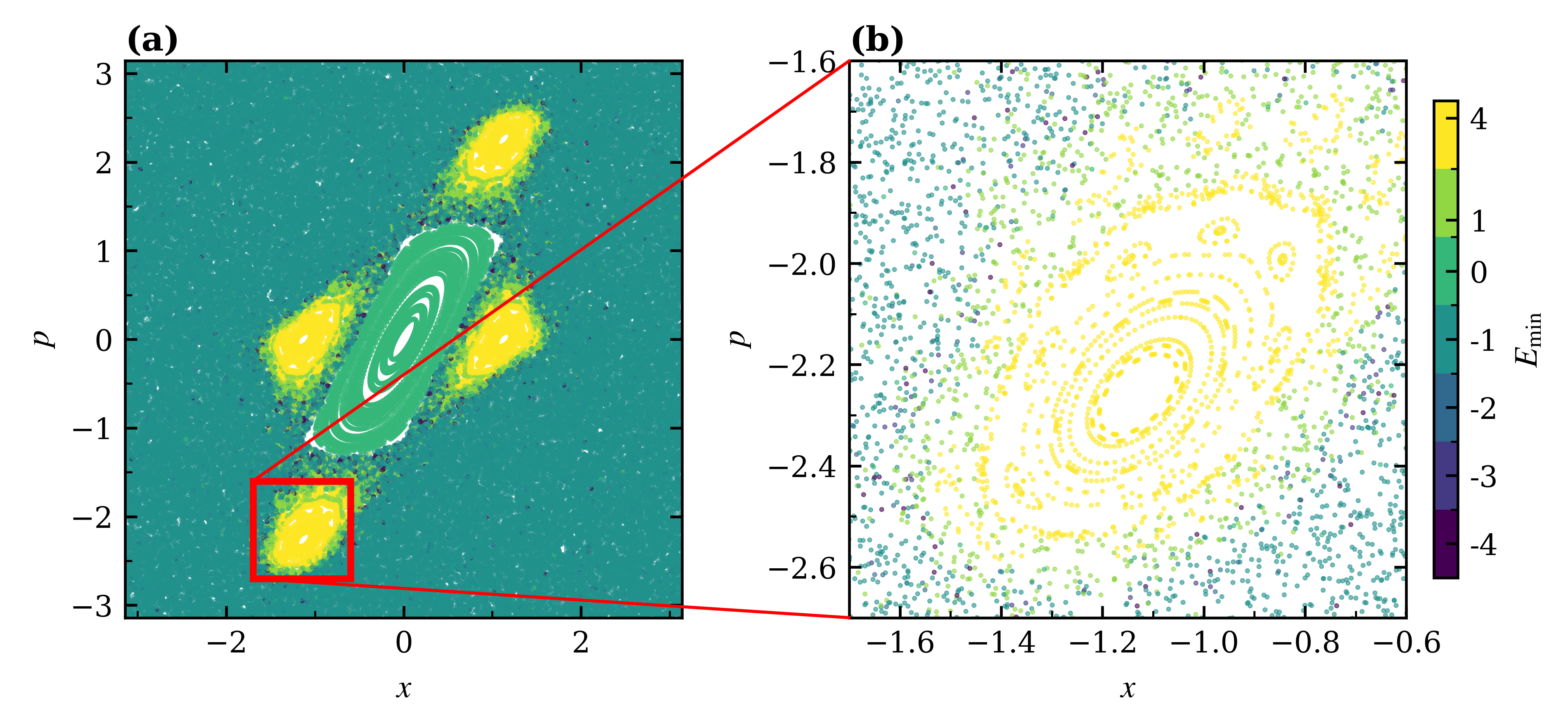}
  \caption{(a) Chirikov standard map phase-space portrait at \(K=2.5\), colored by \(E_{\mathrm{mode}}\), the modal sampled value of \(E(\epsilon)\). (b) Zoom of the region outlined in red.}
  \label{fig:Chirikov2.5EColor}
\end{figure*}

\begin{figure*}[htpb]
  \centering
\includegraphics[width=\textwidth]{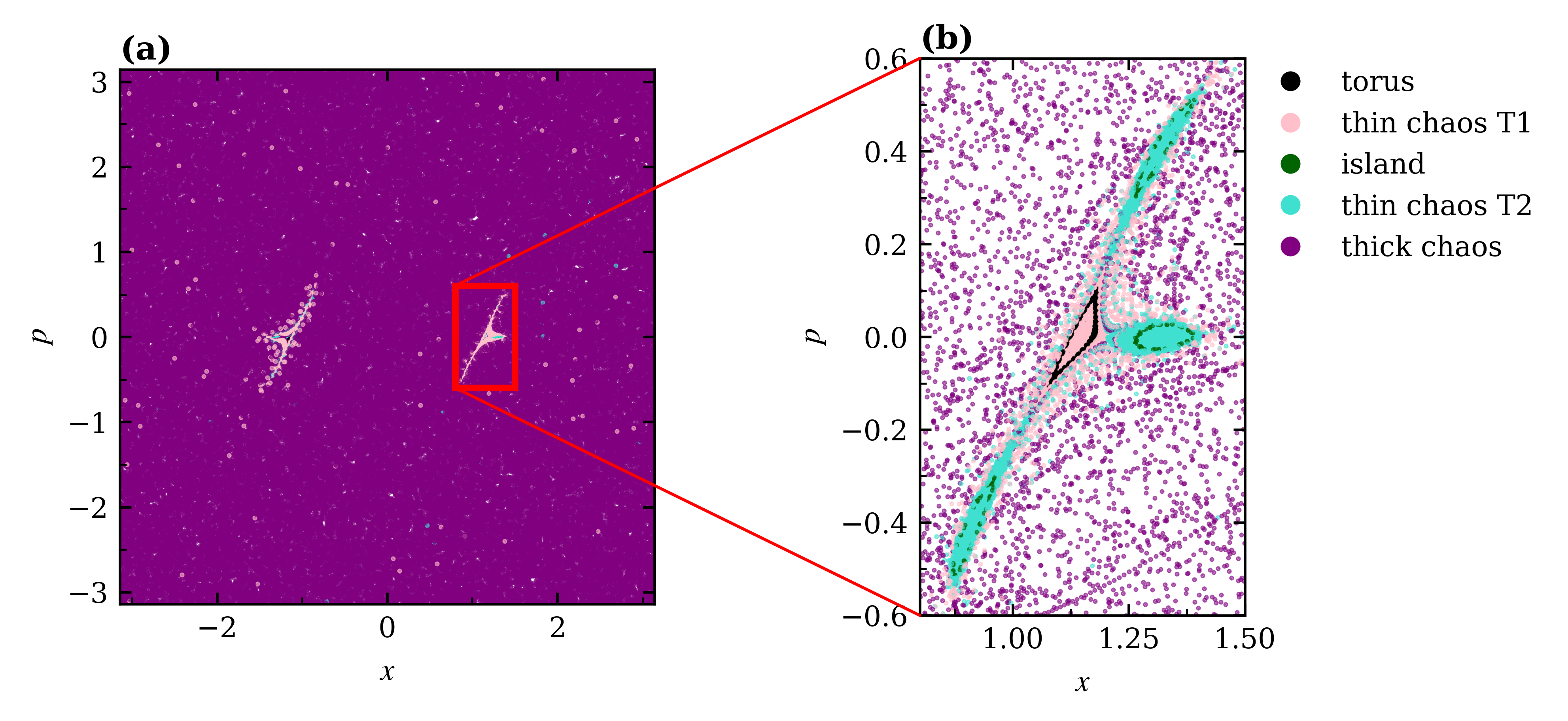}
  \caption{Orbit classification for the Chirikov standard map with $K=6.908745$, $2^{10}$ orbits, $2^7$ points per orbit and all used for classification. (a) Sampled phase-space portrait, orbits are colored according to the Euler--Betti class indicated in the legend. (b) Zoom of the region outlined in red.}
  \label{fig:Chirikov6.9ClassColor}
\end{figure*}

\begin{figure*}[htpb]
  \centering
\includegraphics[width=\textwidth]{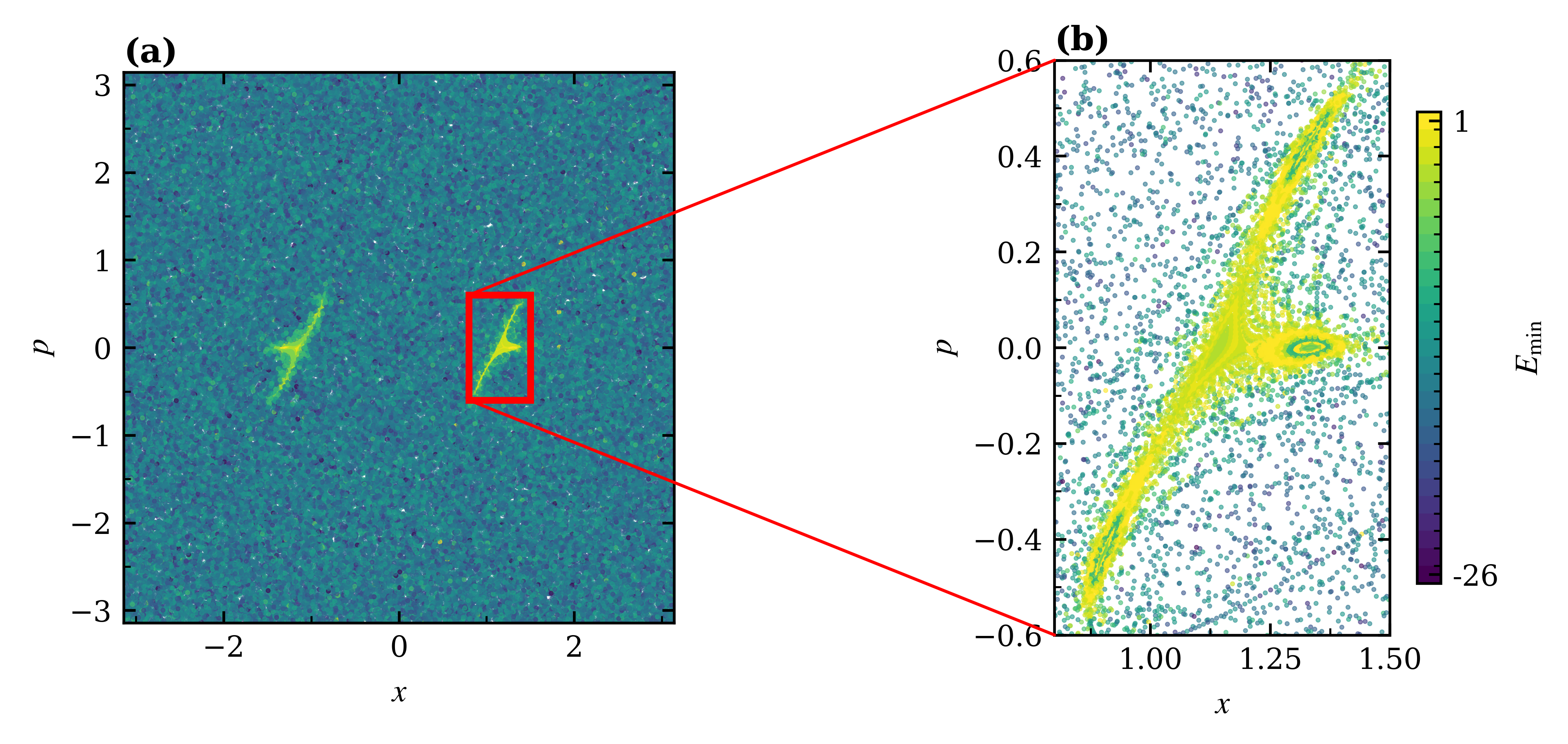}
  \caption{(a) Chirikov standard-map phase-space portrait at \(K=6.908745\), colored by \(E_{\min}\), the minimum sampled value of \(E(\epsilon)\). (b) Zoom of the region outlined in red.}
  \label{fig:Chirikov6.9EColor}
\end{figure*}

We consider $K\in\{0.9, 1.6, 2.5, 6.908745\}$. For the first tree values, the window $[-3,3]^2$ is divided into equal Cartesian cells, with one IC sampled uniformly from each cell. For $K=6.908745$, ICs are sampled uniformly along \(x_0\in[-2.5,2.5]\). The orbit classification results are shown in Figs.~\ref{fig:ChirikovAClassColor},\ref{fig:Chirikov1.6ClassColor},\ref{fig:Chirikov2.5ClassColor},\ref{fig:Chirikov6.9ClassColor}, together with the corresponding \(E_{\mathrm{mode}}\) or \(E_{\min}\) indicator plots in Figs.~\ref{fig:ChirikovBEColor},\ref{fig:Chirikov1.6EColor},\ref{fig:Chirikov2.5EColor},\ref{fig:Chirikov6.9EColor}. These $K$ values 
were chosen both to sample qualitatively distinct phase-space regimes and to facilitate comparison with
representative cases studied in Ref.~\cite{Borin} and in the Hamiltonian-chaos
literature~\cite{DynamicalTraps2002,Zaslavsky2007HamiltonianChaos,Mirella2019}.

Across these representative values of $K$, the Euler--Betti classifier reproduces the
expected progression of the phase-space structure in the standard map. For smaller $K$,
the classified orbits are dominated by regular invariant curves and island chains, while
for intermediate and larger $K$ increasingly broad chaotic layers appear. In particular,
the case $K=6.908745$ highlights the ability of the classifier to detect thin chaotic
regions associated with sticky motion around hierarchical island structures, which are
difficult to characterize from short orbit segments alone. These results indicate that
the Euler--Betti summaries capture the main topological transition from regular to mixed
and strongly chaotic dynamics in the standard map.

\subsection{\label{sec:Harper}Kicked Harper map}

We next consider the kicked Harper map on the unit torus, written in the same
$(x_n,p_n)$ notation as for the Chirikov standard map:
\begin{align}
p_{n+1} &= p_n - 2K \sin(2\pi x_n) \pmod{1},\\
x_{n+1} &= x_n + 2K \sin(2\pi p_{n+1}) \pmod{1}.
\end{align}
As in the previous subsection, $K$ controls the nonlinearity and the
resulting phase-space organization. In the charged-particle interpretation
discussed in Ref.~\cite{Zaslavsky2007HamiltonianChaos}, the kicked Harper
map is associated with magnetic-field effects, in contrast to the
electric-field interpretation of the standard map.

\begin{figure*}[htpb]
  \centering
\includegraphics[width=\textwidth]{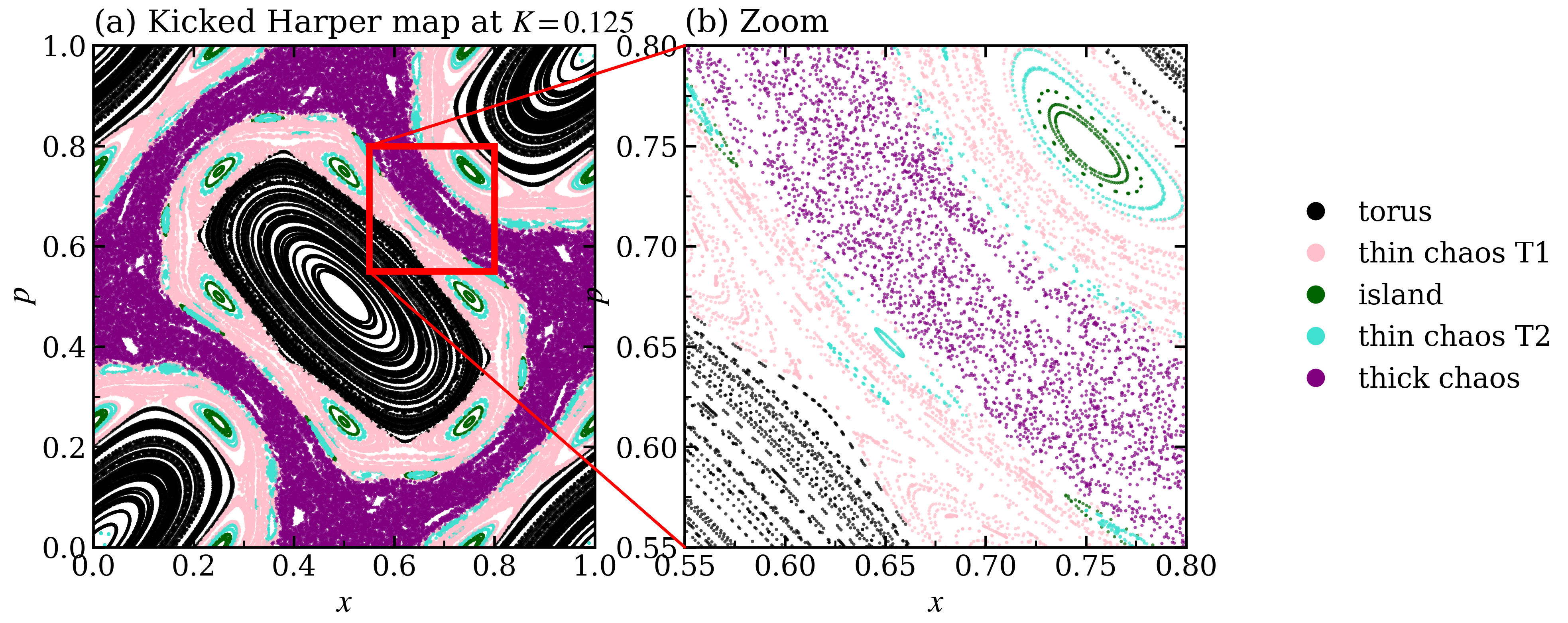}
  \caption{Orbit classification for the kicked Harper map with $K=0.125$, $256$ orbits, $512$ points per orbit and $350$ points used for classification. (a) Sampled phase-space portrait, orbits are colored according to the Euler--Betti class indicated in the legend. (b) Zoom of the region outlined in red.}
  \label{fig:HarperClassColor}
\end{figure*}

\begin{figure*}[htpb]
  \centering
\includegraphics[width=\textwidth]{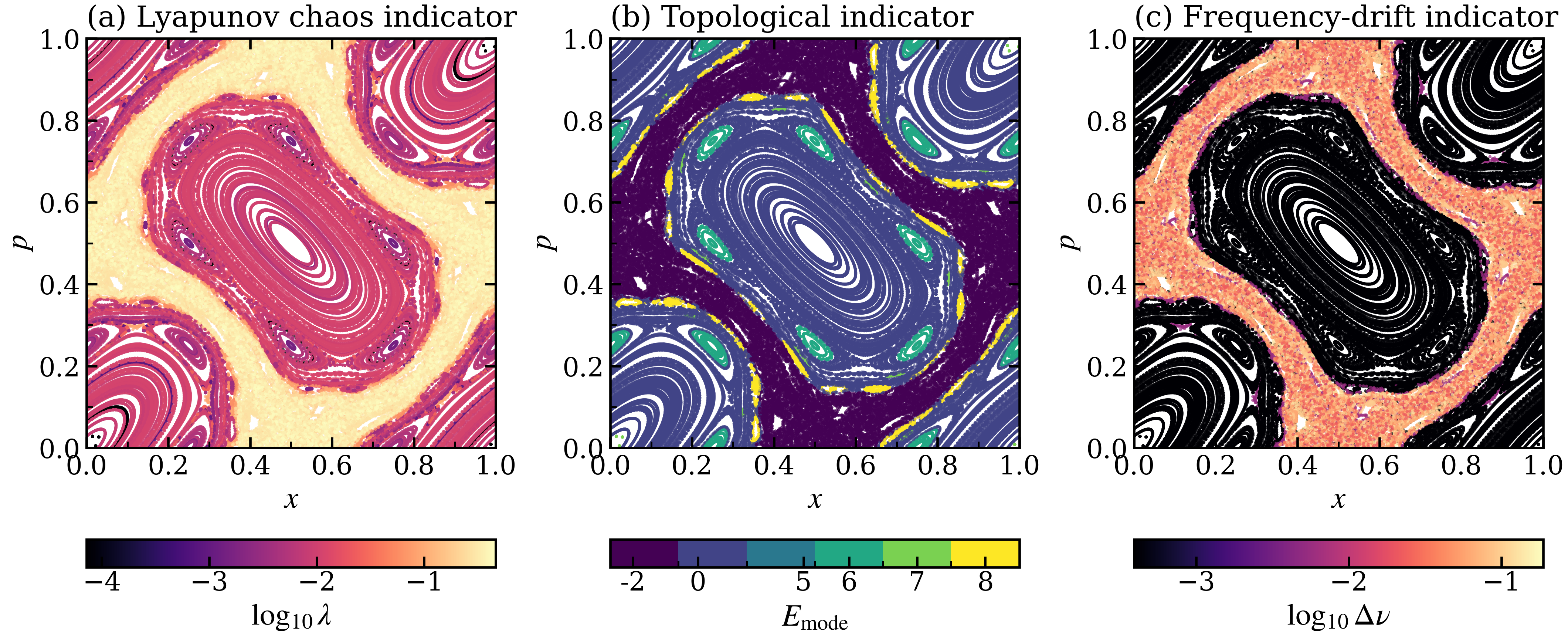}
  \caption{Comparison of (a) the Lyapunov indicator $\log_{10}\lambda$, (b) The topological indicator based on the modal sampled value of $E(\epsilon)$, and (c) a frequency-drift indicator $\log_{10} \Delta \nu$ for the kicked Harper map at $K=0.125$. Persistent homology is computed from 350 points sampled from
each 512-point orbit, whereas
the other two diagnostics use the full 512-point trajectories.}
  \label{fig:ChaosIndicators}
\end{figure*}

The kicked Harper map provides a second 2D benchmark for the
Euler--Betti classifier and complements the standard-map analysis by exhibiting a
different arrangement of regular and chaotic structures on the torus. We assess the classifier using
$K=0.125$, the unit square \([0,1)^2\) is divided into a \(16\times16\) Cartesian partition, with one IC sampled uniformly from each cell. The resulting orbit classification is shown in Fig.~\ref{fig:HarperClassColor}, while Fig.~\ref{fig:ChaosIndicators} compares the PH classification with two established diagnostics: the Lyapunov exponent and frequency map analysis. This allows us to test whether
the PH-based classification is consistent with standard indicators of regular and
chaotic dynamics. 

A notable feature of the kicked-Harper results is that the PH
classification resolves finer phase-space structure even though it is constructed from
shorter orbit segments. In the computations shown here, the Euler--Betti classifier uses
only 350 sampled intersections, whereas both the Lyapunov exponent and frequency map
analysis are computed from 512 iterates. Despite this shorter input, the PH-based
classification reveals sharper separation between regular regions, island chains, and thin
or thick chaotic layers. In this sense, the method appears particularly effective as a
finite-data diagnostic: it retains detailed geometric information from short orbit
segments, while the classical indicators considered here provide a coarser picture of the
same phase-space organization.

This should be contrasted with the Hurst-exponent-based approach of
Ref.~\cite{Borin}, which is also designed to perform well on short data and is
particularly informative about sticky behavior. However, while that method provides a
useful dynamical trap indicator, it does not directly recover the loop and connected-component
structure of the orbit. The Euler--Betti approach therefore complements such chaos indicators by adding an explicit topological description of the orbit.

\subsection{\label{sec:Henon}H\'enon--Heiles map}

We finally consider the H\'enon--Heiles Hamiltonian,
\begin{equation}
H(x,y,p_x,p_y)=\frac{1}{2}(p_x^2+p_y^2)+V(x,y),
\end{equation}
with potential
\begin{equation}
V(x,y)=\frac{1}{2}(x^2+y^2)+x^2y-\frac{1}{3}y^3.
\end{equation}

To apply the 2D Euler-Betti classifier, we construct a Poincar\'e section
with 
$x=0$, $p_x>0$,
and record the section points in coordinates $(y,p_y)$. We study the classical energy values
$H\in\{1/12,1/8,1/6\}$~\cite{HenonHeiles1964ThirdIntegral}, 144 ICs are chosen on this section by sampling admissible
$(y_0,p_{y,0})$ values and recovering $p_{x,0}$ from the energy
constraint:
\begin{equation}
p_{x,0}=\sqrt{2\bigl(H-V(0,y_0)\bigr)-p_{y,0}^2}.
\end{equation}

To facilitate comparison across the $3$ distinct values of $H$, the same set of admissible $(y_0,p_{y,0})$
seeds is used during numerical implementation. Trajectories are integrated with the symplectic St\"ormer--Verlet scheme~\cite{Hairer,CompPhys} using time step
$\Delta t=0.02$. For each IC, we evolve the orbit for $7.5\times10^4$
steps and record intersections with the section whenever the trajectory crosses
$x=0$ with $p_x>0$. PH is then computed only on the resulting
Poincar\'e section, not on the full trajectory.

\begin{figure*}[htpb]
  \centering
\includegraphics[width=\textwidth]{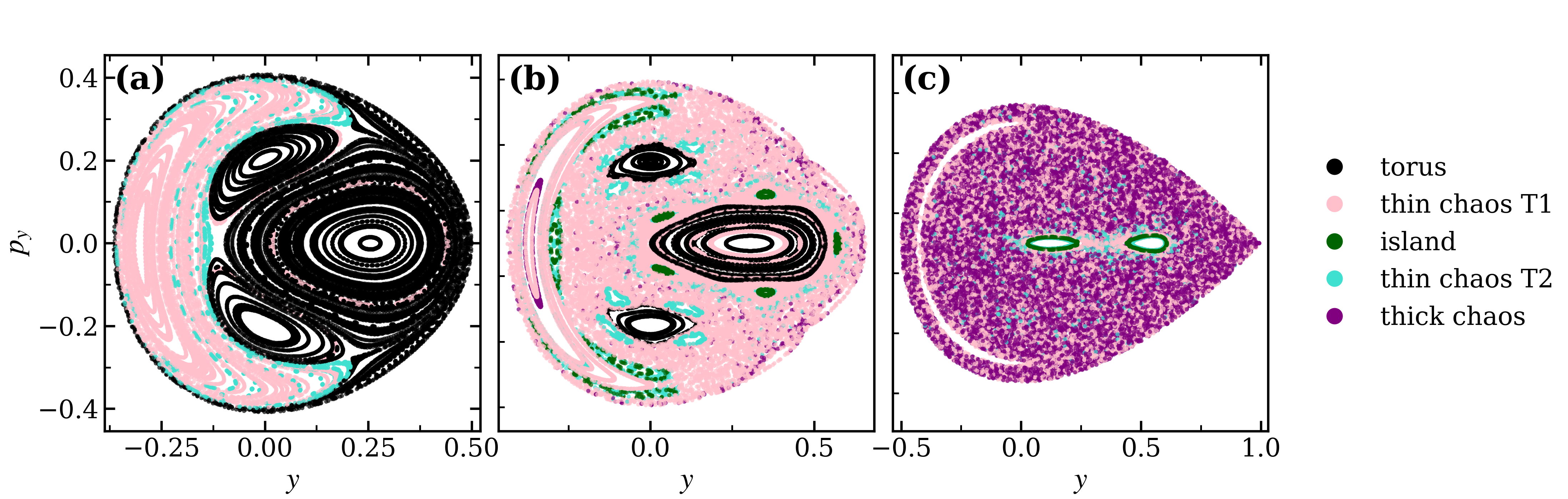}
  \caption{Poincar\'e sections of the H\'enon--Heiles system at energies
(a) $H=1/12$, (b) $1/8$, and (c) $1/6$, colored by the 2D 
Euler--Betti classification. As the energy
increases, the phase space evolves from predominantly torus labels, then a mixture of labels including island structure to broader chaotic regions, with thick chaos
becoming dominant at the highest energy.}
  \label{fig:EnergyComparison}
\end{figure*}

\begin{figure*}[htbp]
  \centering
  \includegraphics[width=\textwidth]{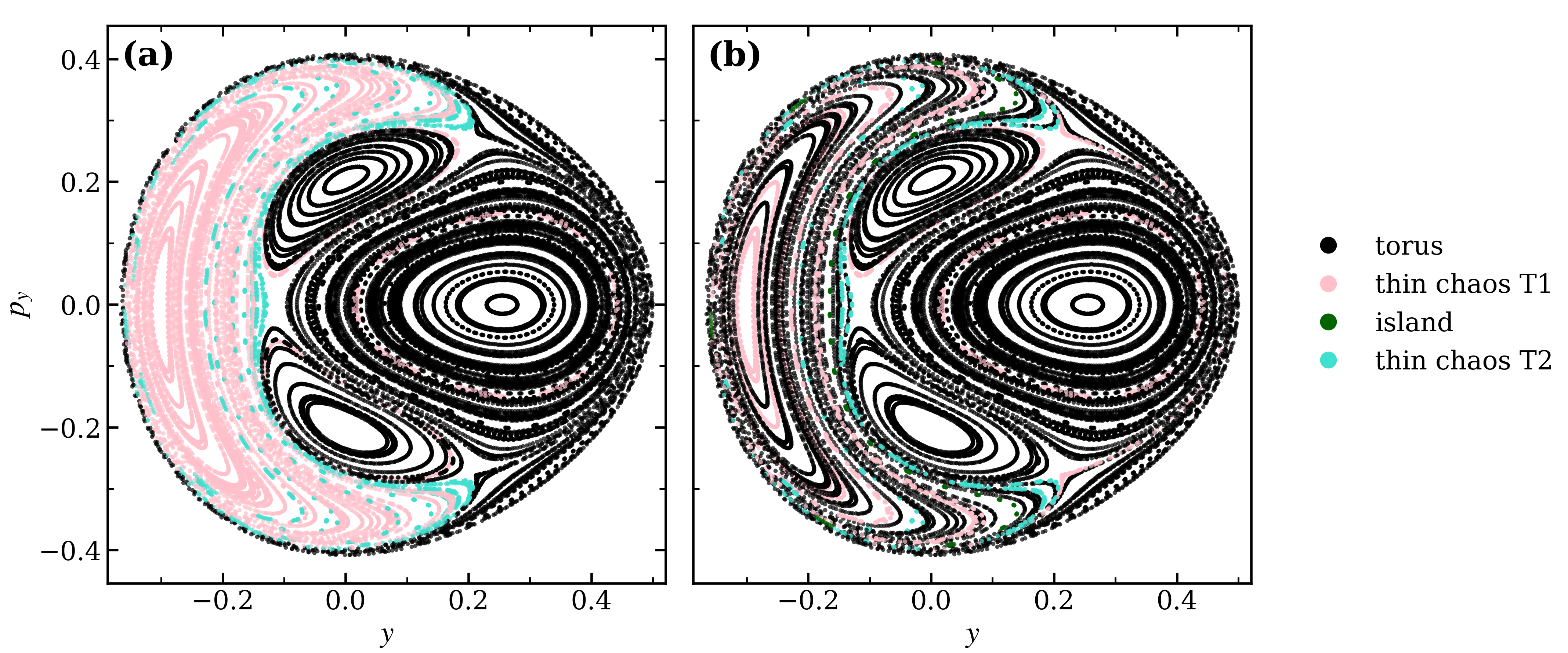}
  \caption{Orbit classifications for the H\'enon--Heiles Poincar\'e section at $H=1/12$. Persistent homology is computed from (a) the intrinsic section $(y,p_y)$, $x=0$, $p_x>0$ and (b) from $(p_x,y,p_y)$.}
  \label{fig:HenonH2d3d}
\end{figure*}

\begin{figure*}[htbp]
  \centering
\includegraphics[width=\textwidth]{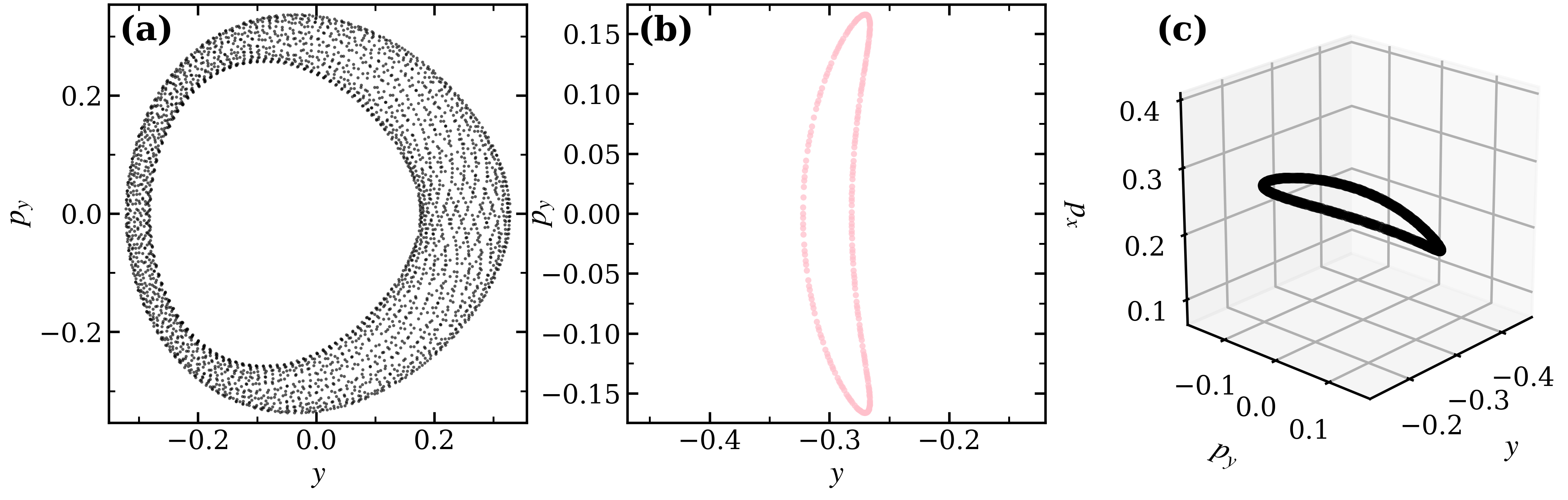}
  \caption{At energy
$H=1/12$, an orbit is initialized at $(x_0,y_0,p_{x,0},p_{y,0})=(0,-0.296860,0.210432,-0.129684)$.
(a) Projection after $7.5\times10^4$
integration steps onto the $(y,p_y)$ plane.
(b) Poincar\'e section $(y,p_y)$, $x=0$, $p_x>0$ with $244$ section intersections.
(c) The same points as in (b) embedded in $(p_x,y,p_y)$. Orbit is classified as (b) thin chaos T1 and (c) torus.}
  \label{fig:BananaLoop}
\end{figure*}

Figure~\ref{fig:EnergyComparison} show the classified Poincar\'e sections for the three energies considered
here. Overall, the Euler--Betti classifier captures the expected mixture of regular
and chaotic orbit families across the energy range. Next, Fig.~\ref{fig:HenonH2d3d} compares two PH classifications at $H=1/2$. The first is the original 2D section and the second is a 3D embedding $(p_x,y,p_y)$ of the same section
points. Both recover the main loop structures, but the
3D embedding more faithfully classifies thin loops that project to banana-shaped
loops in $(y,p_y)$, thereby reducing reducing ambiguities caused by projection. A representative example of this shown in Fig.~\ref{fig:BananaLoop}, where a thin loop with a banana-shaped
appearance in the projected section is misclassified.

This illustrates a known limitation of Vietoris--Rips PH
for thin or poorly separated structures in a Euclidean representation,
as also illustrated in Fig.~3 of Ref.~\cite{Eryilmaz2026}:
points lying on opposite sides of a narrow loop may become connected at
small filtration scales, causing the resulting simplicial complex to fill
the loop before its intended one-cycle is recovered robustly. In the present setting, embedding the section points in
$(p_x,y,p_y)$ improves the geometric separation and leads to a more faithful topological
classification. Possible alternatives for improving the topological analysis of
thin-loop structures are ellipsoid complexes~\cite{Sara2024Ellipsoids,Sara2026ellipsoids,Eryilmaz2026}
or a modified metric in the Vietoris--Rips
construction~\cite{Helene2026}.

\subsection{\label{sec:CoupledChirikov}Coupled Chirikov standard map}

Finally, to test the intrinsic indicators directly in four dimensions,
without reducing the dynamics to a 2D Poincar\'e section
for the Euler--Betti classifier as in Sec.~\ref{sec:Henon}, we consider
a coupled Chirikov standard map written in transverse beam-physics
notation $z=(x,a,y,b)$, where $a$ and $b$ are the scaled momenta conjugate to
$x$ and $y$, respectively. In the minus-sign kick--drift convention used throughout this work, the map is
\begin{align}
a_{n+1}
&=
a_n
-
K_x\sin(x_n)
-
\xi\sin(x_n+y_n),
\label{eq:coupled_chirikov_first}\\
b_{n+1}
&=
b_n
-
K_y\sin(y_n)
-
\xi\sin(x_n+y_n),
\\
x_{n+1}
&=
x_n+a_{n+1},
\\
y_{n+1}
&=
y_n+b_{n+1},
\label{eq:coupled_chirikov_last}
\end{align}
with all variables taken modulo \(2\pi\). Here \(K_x\) and \(K_y\) control the nonlinearities in the two uncoupled standard maps, while \(\xi\) controls the coupling between the two degrees of freedom. When \(\xi=0\), the map reduces to the direct product of two uncoupled Chirikov standard maps in the \((x,a)\) and \((y,b)\) planes. This form is equivalent to the drift--kick convention used in \cite{Lange2014GlobalTori4D} after a one-step shift of the position variables and a sign change of the nonlinear parameters. We use the above form because it matches the sign convention of the 2D Chirikov map used in Sec.~\ref{sec:Chirikov} and the notation used later for beam-physics maps.

\begin{figure*}[htpb]
  \centering
\includegraphics[width=\textwidth]{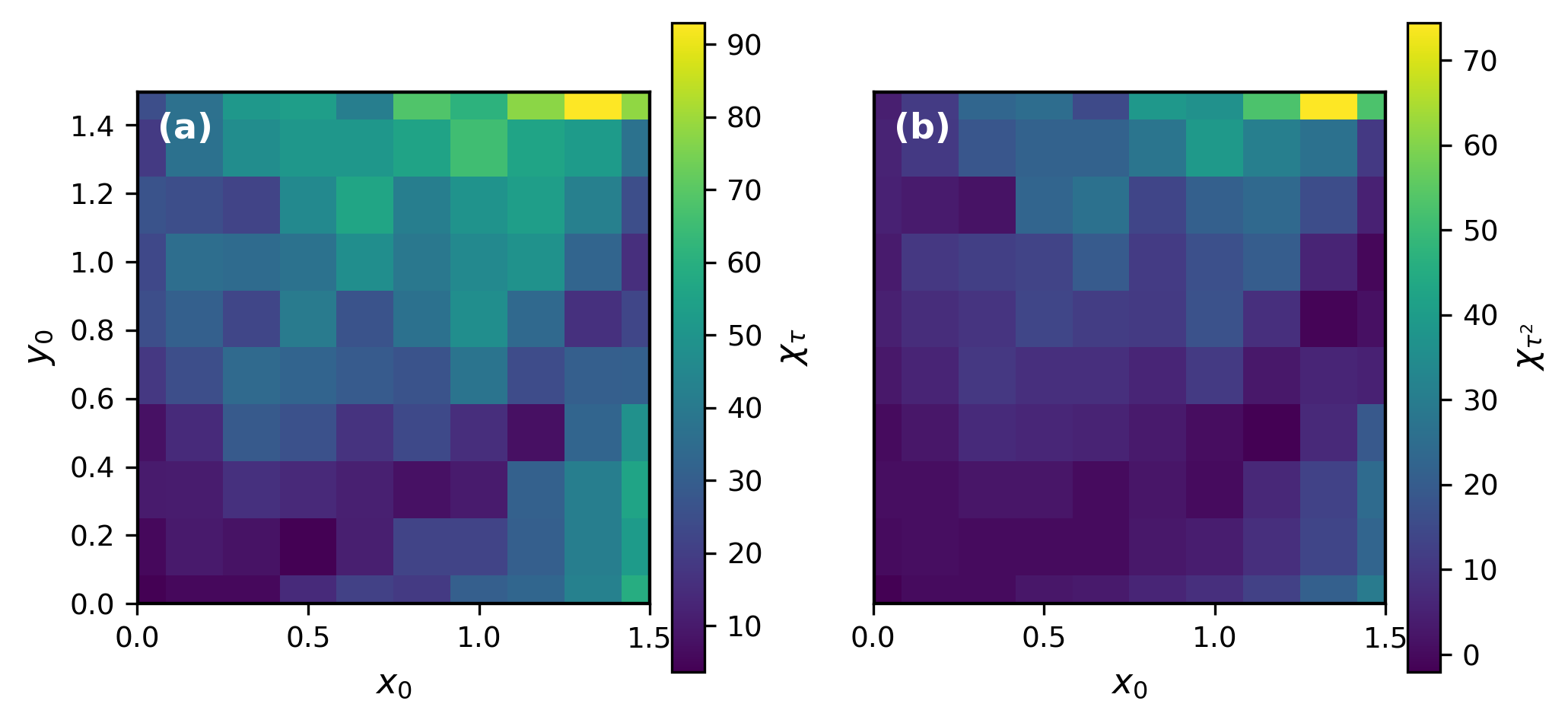}
  \caption{Intrinsic topological chaos indicators on the $10\times10$ grid of
initial conditions for the coupled Chirikov map with
$K_x=1.2$, $K_y=0.9$, and $\xi=0.1$. (a)-(b) Indicators
$\chi_{\tau}$ and $\chi_{\tau^2}$. Both are computed from
the first 150 iterations of each orbit.}
\label{fig:ChaosMetrics}
\end{figure*}

For the calculations below, we use
$K_x=1.2$, $K_y=0.9$, and $\xi=0.1$. We sample a uniform $10\times10$ Cartesian grid over
$(x_0,y_0)\in[0,1.5]^2$, with $a_0=b_0=0$. The origin is excluded from
the subsequent ranking.
For each initial condition, the map is iterated 150 times, and
PH is computed from the resulting 150 orbit points
using the coordinate-wise periodic embedding specified at the beginning
of Sec.~\ref{sec:Validation}. A Vietoris--Rips filtration is constructed, and PH is computed
through homological dimension two. After excluding the unique $H_0$ class, whose death time is infinite, the lifespan sums $\tau^p(B_n)$ are computed in each dimension for
$p\in\{1,2\}$. The resulting intrinsic indicators
$\chi_{\tau}$ and $\chi_{\tau^2}$, defined in
Sec.~\ref{sec:Intrinsic}, are displayed in
Fig.~\ref{fig:ChaosMetrics}.

\begin{figure*}[htpb]
  \centering
\includegraphics[width=\textwidth]{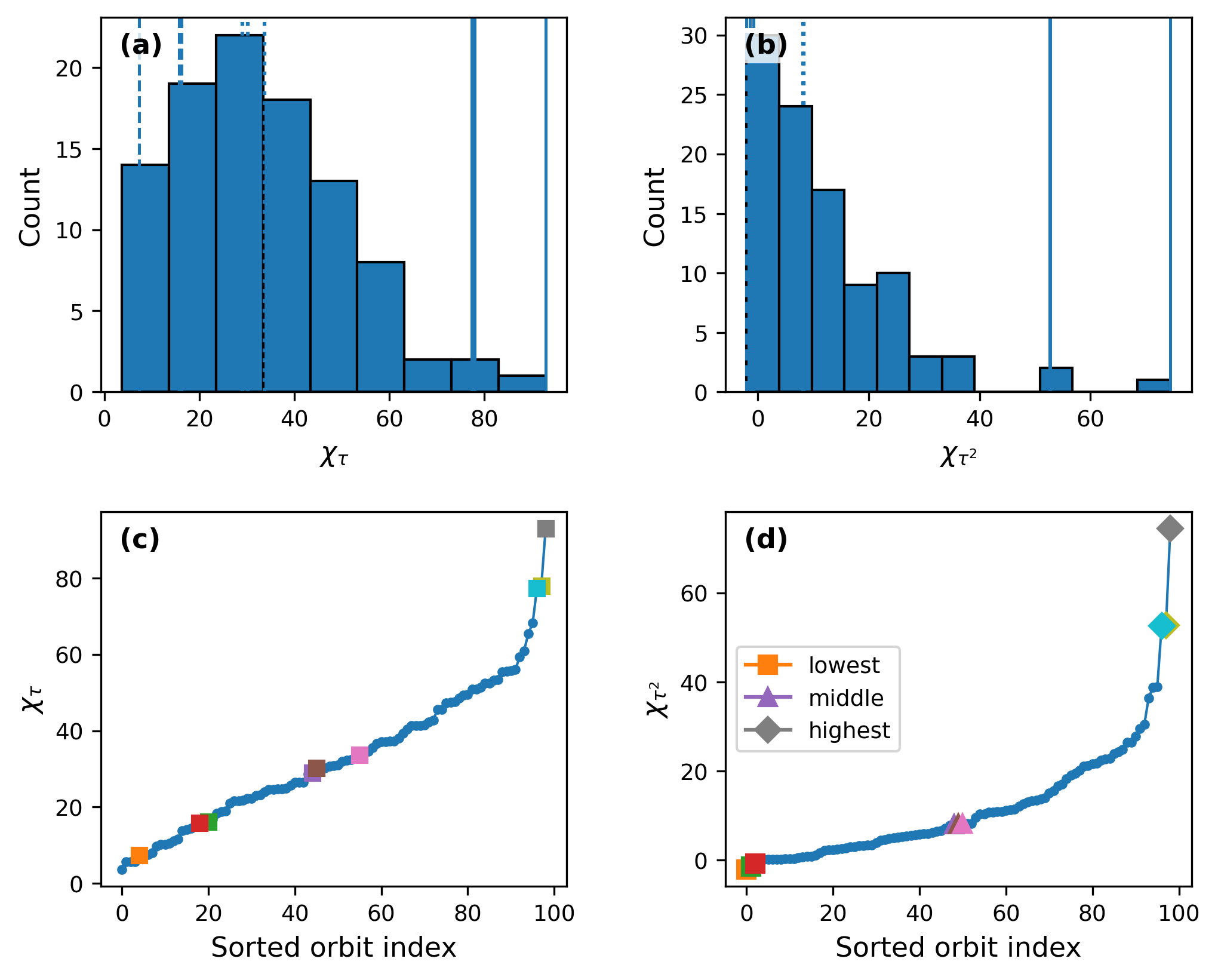}
  \caption{Distributions (a)-(b) and rank-ordered values (c)-(d) of
$\chi_{\tau}$ (left) and $\chi_{\tau^2}$ (right) for the 99 origin-excluded initial
conditions of the coupled Chirikov map. Both
rankings are computed from the first 150 iterations. The markers identify the three
lowest-, middle-, and highest-ranked initial conditions.}
  \label{fig:Ranking}
\end{figure*}

The ICs are ranked separately according to each indicator.
Figure~\ref{fig:Ranking} highlights the three lowest-, three middle-,
and three highest-ranked ICs and shows the corresponding distributions. To examine the extended dynamics of the selected orbits, we iterate
each one for 5000 steps and consider projections onto $(x,a,y)$,
$(x,a)$, $(y,b)$, and $(x,y)$. The extended trajectories exhibit the qualitative ordering anticipated
from the intrinsic indicators.

\begin{figure*}[htpb]
  \centering
\includegraphics[width=\textwidth]{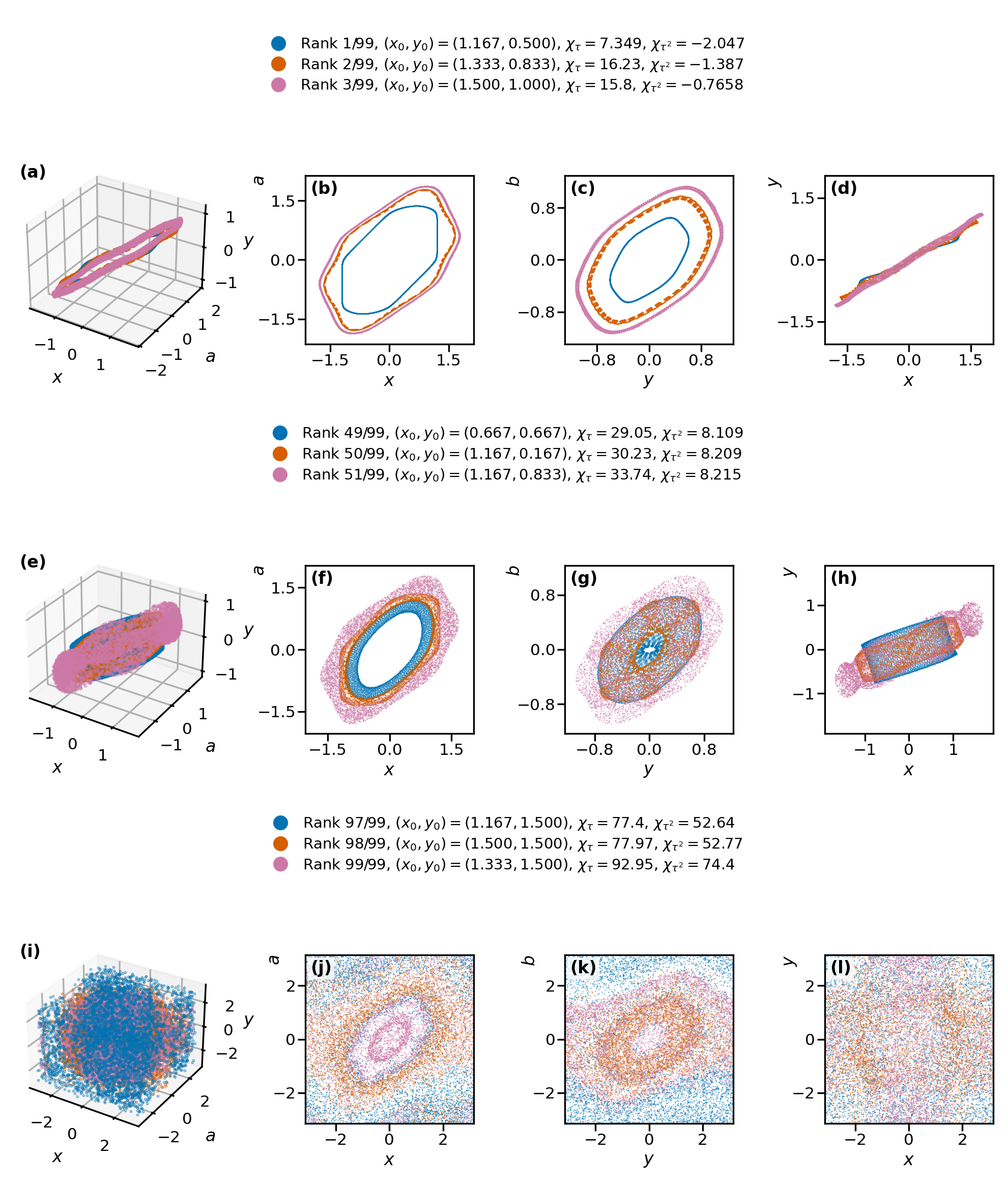}
  \caption{Long-orbit projections for the coupled Chirikov map with
$K_x=1.2$, $K_y=0.9$, and $\xi=0.1$.(a)-(d) Lowest-, (e)-(h) middled-, (i)-(l) highest-ranked initial conditions according to $\chi_{\tau^2}$. The ranking is computed from
the first 150 iterations, while the displayed trajectories contain
5000 iterations.  From left to right, each row show the
three-dimensional projection $(x,a,y)$ and the two-dimensional
projections $(x,a)$, $(y,b)$, and $(x,y)$. Colors distinguish the three
initial conditions.}
  \label{fig:LongerProjections}
\end{figure*}

The three lowest-ranked orbits shown in (a)-(d) of
Fig.~\ref{fig:LongerProjections} remain confined to compact, coherent structures, with smooth and well-organized projections characteristic of regular or quasi-periodic motion.

The middle-ranked orbits shown in (a)-(h) of Fig.~\ref{fig:LongerProjections} occupy
broader and visibly thicker regions of phase space. Although remnants of organized loop-like geometry remain, the trajectories show increased transverse spreading, multiple strands, and a partial loss of the smooth invariant-curve structure seen at low rank.

Finally, the three highest-ranked orbits shown in (i)-(l) of Fig.~\ref{fig:LongerProjections} display the greatest geometric dispersion: their points extend over broad portions of the projected regions and no longer remain confined to a single thin or smoothly organized structure, which is consistent with strongly chaotic motion.

The progression from compact coherent sets, through broadened intermediate structures, to widely dispersed orbit clouds is consistent with the increase of both $\chi_{\tau}$ and $\chi_{\tau^2}$. The separation is especially pronounced for $\chi_{\tau^2}$ because squaring the persistence lifespans gives greater relative weight to long-lived topological features. 

\section{\label{sec:IOTA}Beam-physics case study: IOTA Lattice}

\begin{figure*}[htbp]
  \centering
\includegraphics[width=\textwidth]{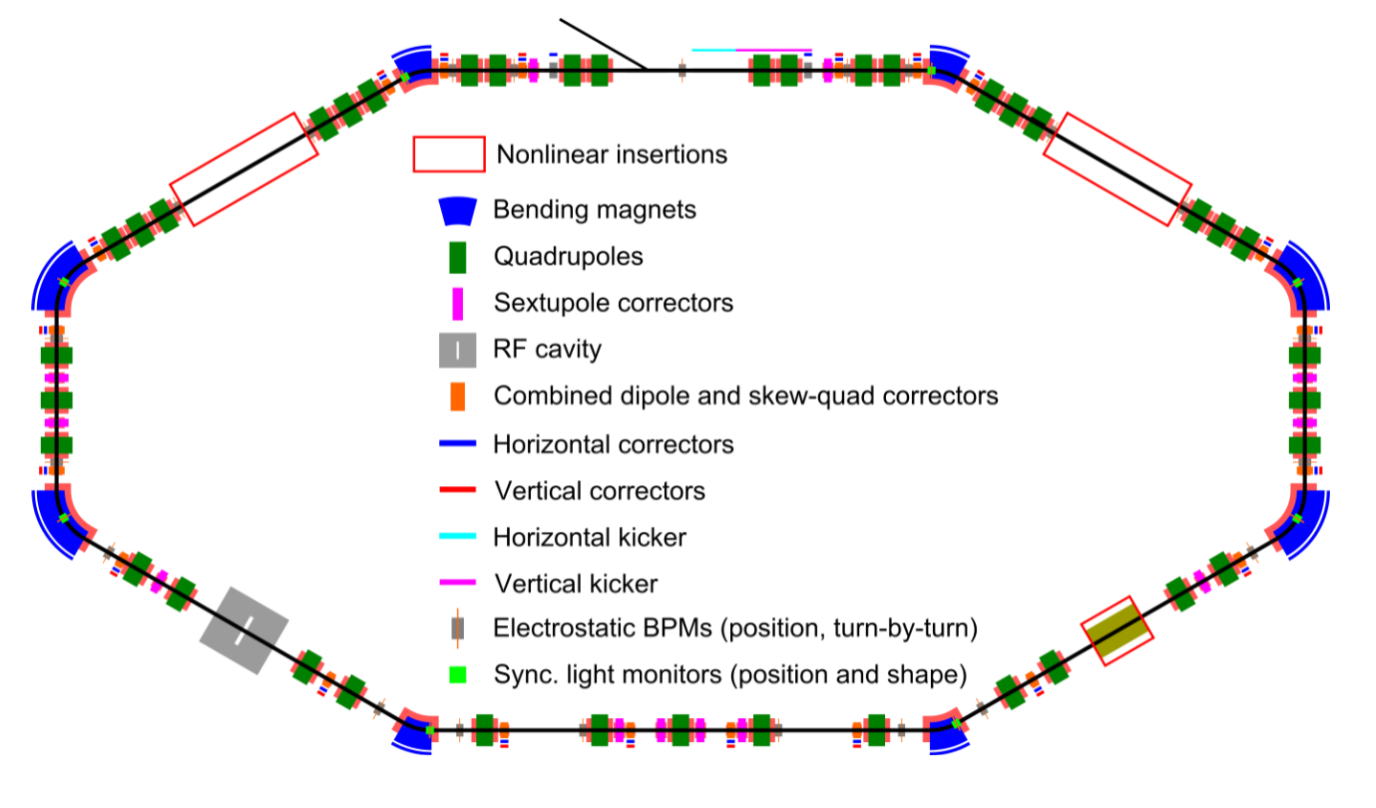}
  \caption{Layout of the Integrable Optics Test Accelerator (IOTA) ring. Adapted from Ref.~\cite{Antipov2017IOTA} under CC BY 3.0 (https://creativecommons.org/licenses/by/3.0/), with nonlinear magnetic
inserts added.}
  \label{fig:IOTA}
\end{figure*}

The IOTA ring, shown in Fig.~\ref{fig:IOTA}, is an operating beam physics test facility at Fermi National Accelerator Laboratory. It serves as a testbed for novel ideas in accelerator science and a blueprint for future accelerator systems. Specifically, it has an octagonal geometry, with eight bending magnets located at the ends of the straight sections and a perimeter of approximately \(40~\unit{m}\). IOTA provides a natural test case for the present method because its
nonlinear inserts are designed to generate strongly nonlinear but
controlled transverse beam dynamics. In particular, the Danilov--Nagaitsev (DN) nonlinear potential gives rise to lattices whose dynamics are relevant for integrable-optics studies in beam physics~\cite{Antipov2017IOTA}.

\subsection{\label{sec:Normalization}One-turn map, normal form transformation, and persistent homology workflow}

A reference particle traveling around the ring provides the baseline against which the motion of other particles is measured. Relative to this particle, a six-dimensional beam-dynamics description contains horizontal, vertical, and longitudinal phase-space pairs. The longitudinal pair describes differences in arrival time and energy relative to the reference particle. The present work restricts attention to transverse dynamics at the reference energy. At a fixed observation point around the ring, the transverse phase-space state is denoted by $
z=(x,a,y,b)$,
where $x$ and $y$ are the horizontal and vertical displacements from the reference trajectory, and $a=p_x/p_0$ and $b=p_y/p_0$ are the corresponding scaled transverse momenta, with $p_0$ denoting the reference momentum.

The periodicity of the lattice allows the dynamics to be represented by a one-turn map $\mathcal{M}$ \cite{Berz1999MapMethods}. The high-order truncated Taylor maps used here are generated in COSY INFINITY and
then symplectified using the extended Poincar\'e (EXPO) generating-function
method~\cite{Bela,Bela2004cosy}. We denote the resulting EXPO-symplectified one-turn map by
$\widehat{\mathcal{M}}$. After $N$ turns, the tracked coordinates are
\begin{equation}
z_N=\widehat{\mathcal{M}}^{N}(z_0),
\label{eq:iota_tracking}
\end{equation}
where $\widehat{\mathcal{M}}^{N}$ denotes the $N$-fold composition of
$\widehat{\mathcal{M}}$.
Within the differential algebra framework~\cite{Berz1999MapMethods}, the symplectification can be
carried out to an arbitrary finite truncation order. In the computations
below, all maps are truncated at order~10.
 
Before computing the PH diagnostics, each
orbit is checked for loss in the original $z$ coordinates. A trajectory is marked as lost if $\sqrt{x^2+y^2}>50~\mathrm{mm}$. Only orbits that
survive the sampled short-orbit interval are used for PH; the remaining
ICs are assigned a loss label. For the extrinsic
classification of Sec.~\ref{sec:Extrinsic Classification}, the surviving
orbits are transformed as
$\bar{z}_N=\mathcal{A}\circ\widehat{\mathcal{M}}^{\,N}(z_0)$, where $\mathcal{A}$ is the nonlinear transformation that brings a
symplectic one-turn map $\mathcal {M}$ into normal form~\cite{Berz1993DA_normal_form},
\begin{equation}
    \mathcal{N}
    =
    \mathcal{A}\circ \mathcal{M}\circ \mathcal{A}^{-1}.
\end{equation}

Each transformed 4D orbit data set is projected onto the
two symplectic planes, producing two 2D data sets, and the
classifier of Table~\ref{tab: stability classifier} is applied separately
to them. The resulting combined label is assigned to the original IC $z_0$, so the PH results are displayed in the original
launch coordinates.

For the 4D intrinsic indicators and the
2D flat-beam analysis below, PH is not computed directly in the raw $z$ coordinates. 
Since the coordinates $(x,a,y,b)$ may have different natural scales, raw orbit clouds can be strongly anisotropic, which can introduce artificial 
geometric distortions into Vietoris--Rips PH. We therefore use the linearly
scaled transverse coordinates $\tilde z_N=\mathcal{A}_L\circ\widehat{\mathcal{M}}^N(z_0)$,
where \(\mathcal{A}_L\) is the linear part of  \(\mathcal{A}\).
This linear scaling reduces trivial scale imbalances among the position
and momentum coordinates by placing them on comparable scales. It
therefore provides a more suitable representation of the orbit data for
the Euclidean metric used in the Vietoris--Rips filtration and may improve
the recovery of thin loop-like structures, such as those encountered in
the H\'enon--Heiles example. However, linear scaling alone does not eliminate all possible sources of
topological misclassification.

The IOTA maps considered here have midplane symmetry, so the plane
$y=b=0$ is invariant~\cite{BelaSymmetries}. Trajectories initialized as $z_0=(x_0,a_0,0,0)$ therefore remain in this plane and produce a 2D Poincar\'e section corresponding to a flat-beam configuration. The PH workflow for each particle species below is then:
\begin{enumerate}

\item Guided by the lattice-dependent characteristic phase-space scales, select a feasible $z_0=(x_0,a_0)$ window with $y_0=b_0=0$. Track all ICs in this window for $N_{\mathrm{flat}}$ turns and apply the 2D Euler--Betti classifier to the corresponding scaled data $\tilde{z}_{N_{\mathrm{flat}}}$ to identify the extent of a nested-torus region. The validation results show that a few hundred turns are sufficient.

\item Use the horizontal scale identified in the previous flat-beam analysis,
with an additional margin in both launch coordinates, to define a
transverse grid with $z_0=(x_0,0,y_0,0)$. Choose a short turn number $N_{\mathrm{tr}}$, and for each surviving IC, compute the extrinsic classification using $\bar{z}_{N_{\mathrm{tr}}}$ coordinates.
\item For the same surviving transverse ICs, compute the intrinsic
indicator $\chi_{\tau^2}$ from the full 4D linearly scaled coordinates $\tilde z_{N_{\mathrm{tr}}}$. 
\end{enumerate}

After this, each transverse IC $z_0$ is assigned a categorical extrinsic label and a scalar intrinsic indicator. From the developed homological theory, it is reasonable to expect an origin-connected \(S\times S\) region to lie within a tracking-based dynamic aperture and to be generally associated with lower values of \(\chi_{\tau^2}\). The benchmarks below test these expectations.

To benchmark the short-orbit PH diagnostics, we also perform
long-term tracking of the transverse ICs from step 2 above. At discrete checkpoints, the recorded coordinates
are evaluated using the same loss criterion.
If COSY stops reporting a trajectory before the final checkpoint, the first
missing checkpoint is treated as the observed loss checkpoint. We define
$N_s$ as the first sampled turn at which loss is observed; thus, $N_s$
records an observed checkpoint rather than the exact loss turn. Particles
that survive through the final checkpoint are assigned
$N_s=N_{\max}$, and the survival plots display $\log_{10}(N_s)$.

\subsection{Electron case}

IOTA can be utilized to store various charged particle species. Here we study two especially interesting cases for beam physics, electrons and protons. We begin with a flat electron beam launched in the nonlinear IOTA lattice. The
singularity parameter of the DN nonlinear magnet is fixed at $c=0.03$,
and its strength is set to $t=0.223$, with the
sextupoles and octupoles turned on. The electron reference energy is
$149.49~\mathrm{MeV}$, and the transverse beam widths are $(\sigma_x,\sigma_y)=(0.583095,0.4)~\mathrm{mm}$. In the product
interpretation of Sec.~\ref{sec:4DIdealized}, the flat-beam orbits have classes of the form $\mathcal C\times\mathrm{FP}$, where
$\mathcal C$ is the horizontal orbit class.
 
\begin{figure*}[htpb]
  \centering
\includegraphics[width=\textwidth]{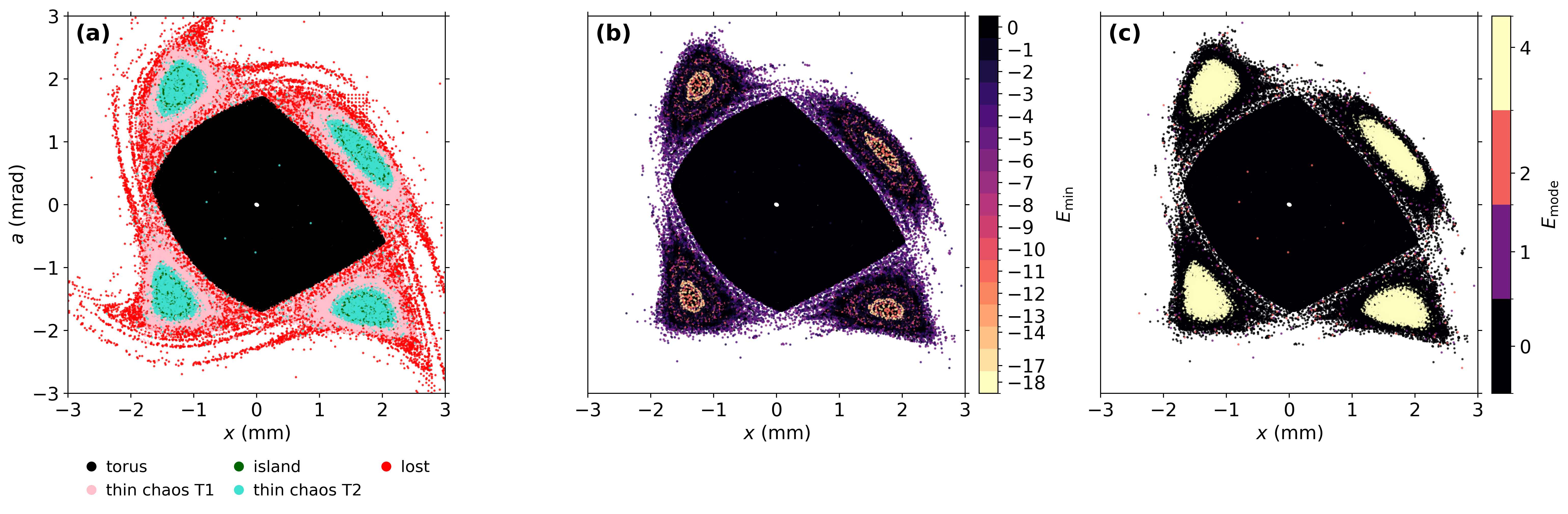}
  \caption{Persistent homology analysis of the electron flat-beam case.
The 961 selected initial conditions form a $31\times31$ uniform grid over a
$1.75~\mathrm{mm}\times1.75~\mathrm{mrad}$ horizontal phase-space
window; the $x_0$ extent corresponds to approximately $3\sigma_x$. PH is computed from 200-turn orbits in $(\tilde{x},\tilde{a})$ and displayed in the raw $(x,a)$ coordinates. (a) Numerical orbit classes obtained from the 2D Euler--Betti classifier, with lost trajectories shown separately. (b) Minimum sampled Euler--Betti value $E_{\min}$ assigned to each orbit. (c) Modal Euler--Betti value $E_{\mathrm{mode}}$ assigned to each orbit.}
  \label{fig:FlatElectron}
\end{figure*}

For the flat-beam scan, we sample 961 ICs on a
$31\times31$ uniform grid in the horizontal launch plane $(x_0,a_0)$.
The grid spans a $1.75~\mathrm{mm}\times1.75~\mathrm{mrad}$ phase-space
window. Each IC is tracked for $N_{\mathrm{flat}}=200$ turns. Figure~\ref{fig:FlatElectron}
shows the resulting Euler--Betti classification together with the
quantities $E_{\min}$ and $E_{\mathrm{mode}}$, and panel (a) shows that the nested-torus region
extends horizontally to approximately
$3\sigma_x\simeq1.75~\mathrm{mm}$. 

Guided by this scale, we next consider
transverse launches on a uniform $(x_0,y_0)$ grid covering a
$4\sigma_x\times4\sigma_y$ region, with zero initial momenta
$a_0=b_0=0$. For each IC, the extrinsic classification and the intrinsic indicator
$\chi_{\tau^2}$ are computed from the
first $N_{\mathrm{tr}}=200$ turns. The transverse grid is also tracked to
$N_{\max}=10^4$ turns, with loss evaluated every 100 turns, to construct
the survival diagnostic $\log_{10}(N_s)$ and obtain a tracking-based DA. 

\begin{figure*}[htpb]
  \centering
\includegraphics[width=\textwidth]{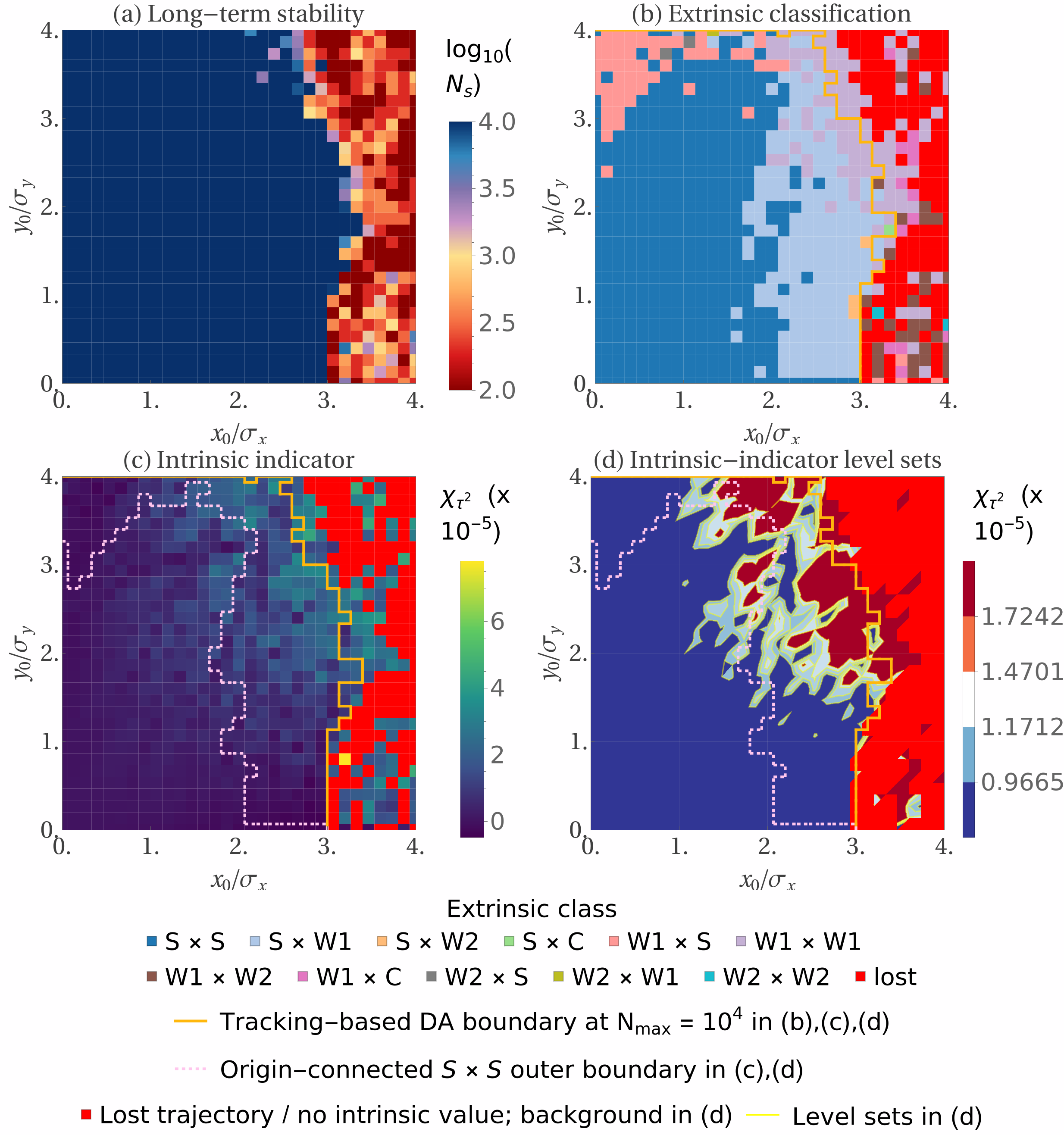}
\caption{Electron transverse-launch stability diagnostics for the
nonlinear IOTA lattice considered here. (a)
Long-term stability index $\log_{10}(N_s)$, initial-condition
grid tracked to $N_{\max}=10^4$ turns, with loss checked every 100 turns. (b) Classification obtained using the nonlinear transformation
$\mathcal{A}$ and classifying the two symplectic-plane projections. (c) Intrinsic indicator $\chi_{\tau^2}$
computed in linearly scaled coordinates, and (d) its linearly interpolated level sets at the 65th, 70th, 75th and 80th percentiles. All PH diagnostics
are computed from the first 200 turns.} 
\label{fig:TransverseElectron}
\end{figure*}

The results are displayed in Fig.~\ref{fig:TransverseElectron}, which shows broad spatial agreement among the long-term
survival map (a), the extrinsic classification (b), and the 
intrinsic indicator (c). In particular, panel (b) confirms that the connected
$\mathrm{S}\times\mathrm{S}$ region containing the reference particle
lies within the region that survives through the full
$10^4$-turn run, and it also generally corresponds to smaller values of
$\chi_{\tau^2}$, see panels (c), (d). Near the tracking-based DA boundary, the extrinsic classification transitions to a mixture of weakly chaotic labels in one or both planes and loss flags, while survival times generally decrease beyond the boundary. Similarly, near the DA, $\chi_{\tau^2}$ becomes more heterogeneous, with a mixture of lower and intermediate values, the largest one being beyond the DA and corresponding to a $\mathrm{W2}\times\mathrm{W2}$ class. These results indicate that the short-orbit PH diagnostics
capture the main structure of the transverse stability boundary.

\subsection{Proton case}

We repeat the previous workflow for the proton configuration, but use a larger maximum turn number $N_{\max}$ to determine the tracking-based DA. Note that while a short term dynamic aperture is a sufficient indicator for the electron case due to the significant synchrotron radiation emitted even at low energies, that is not the case for the protons; in this case, lack of significant radiation at these energies necessitates the knowledge of the long-term dynamic aperture. The proton
reference energy is $2.5~\mathrm{MeV}$, and the transverse beam widths are $(\sigma_x,\sigma_y)=(1.59687,1.09545)~\mathrm{mm}$.
The
corresponding one-turn map is generated in COSY INFINITY and then
symplectified, with the sextupoles and octupoles turned on. The DN
parameters remain $c=0.03$ and $t=0.223$, as in the electron case.

\begin{figure*}[htpb]
  \centering
\includegraphics[width=\textwidth]{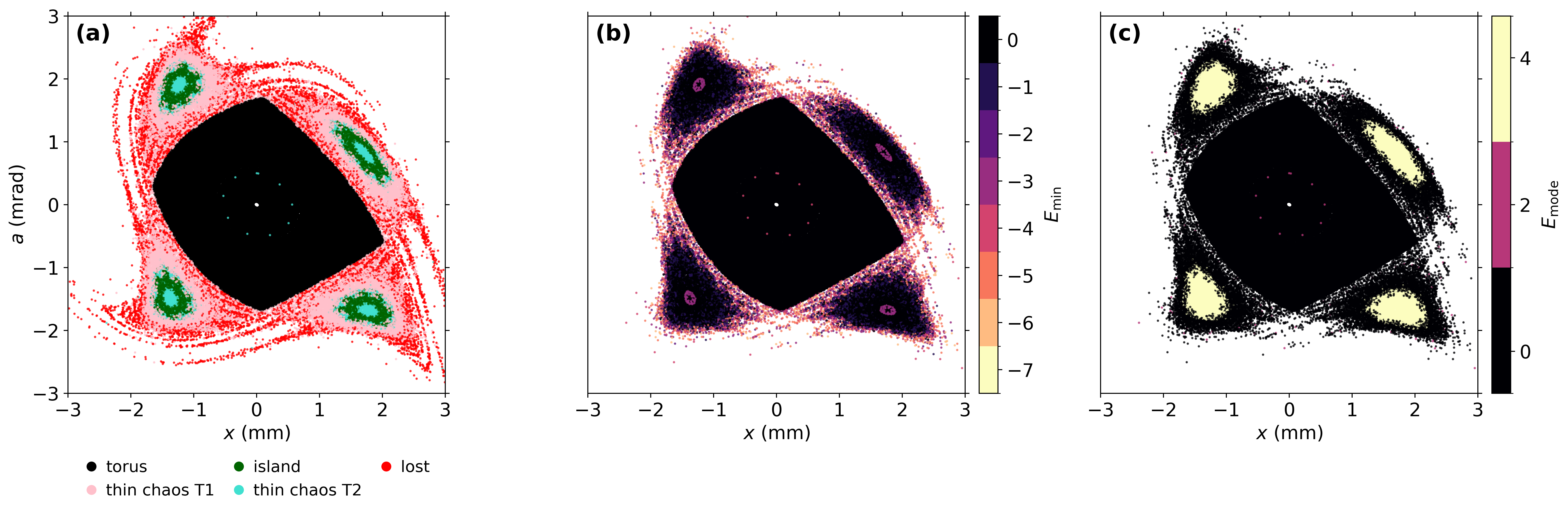}
  \caption{Persistent homology analysis of the proton flat-beam case. The 961 selected initial conditions form a $31\times31$ uniform grid spanning $1.60~\mathrm{mm}\times1.60~\mathrm{mrad}$, with the $x_0$ range corresponding to approximately $1\sigma_x$. PH is computed from 200-turn orbits in $(\tilde{x},\tilde{a})$ and displayed in the raw $(x,a)$ coordinates. (a) Numerical orbit classes obtained from the 2D Euler--Betti classifier, with lost trajectories shown separately. (b) Minimum sampled Euler--Betti value $E_{\min}$ assigned to each orbit. (c) Modal Euler--Betti value $E_{\mathrm{mode}}$ assigned to each orbit.
}
  \label{fig:FlatProton}
\end{figure*}

We sample 961 ICs on a
$31\times31$ uniform grid in the horizontal launch plane $(x_0,a_0)$.
The grid spans a
$1.60~\mathrm{mm}\times1.60~\mathrm{mrad}$ phase-space window. Each
IC is tracked for 200 turns. The results are shown in
Fig.~\ref{fig:FlatProton}, this time, the nested-torus region
extends horizontally to approximately
$1\sigma_x\simeq1.60~\mathrm{mm}$. 

Guided by the horizontal extent of the
nested-torus region in the flat-beam scan, we perform the transverse-launch
analysis on a uniform $(x_0,y_0)$ grid spanning a
$2\sigma_x\times2\sigma_y$ region, with zero initial momenta
$a_0=b_0=0$. The extrinsic
classification and the intrinsic indicator $\chi_{\tau^2}$ are computed from the first 200 turns. For the survival diagnostic, three tracking outputs are combined:
loss is checked every $10^3$ turns in the run to $10^5$ turns,
every $10^4$ turns in the run to $10^7$ turns, and every $10^5$
turns in the run to $N_{\mathrm{max}}=10^8$ turns. 

\begin{figure*}[htpb]
  \centering
\includegraphics[width=\textwidth]{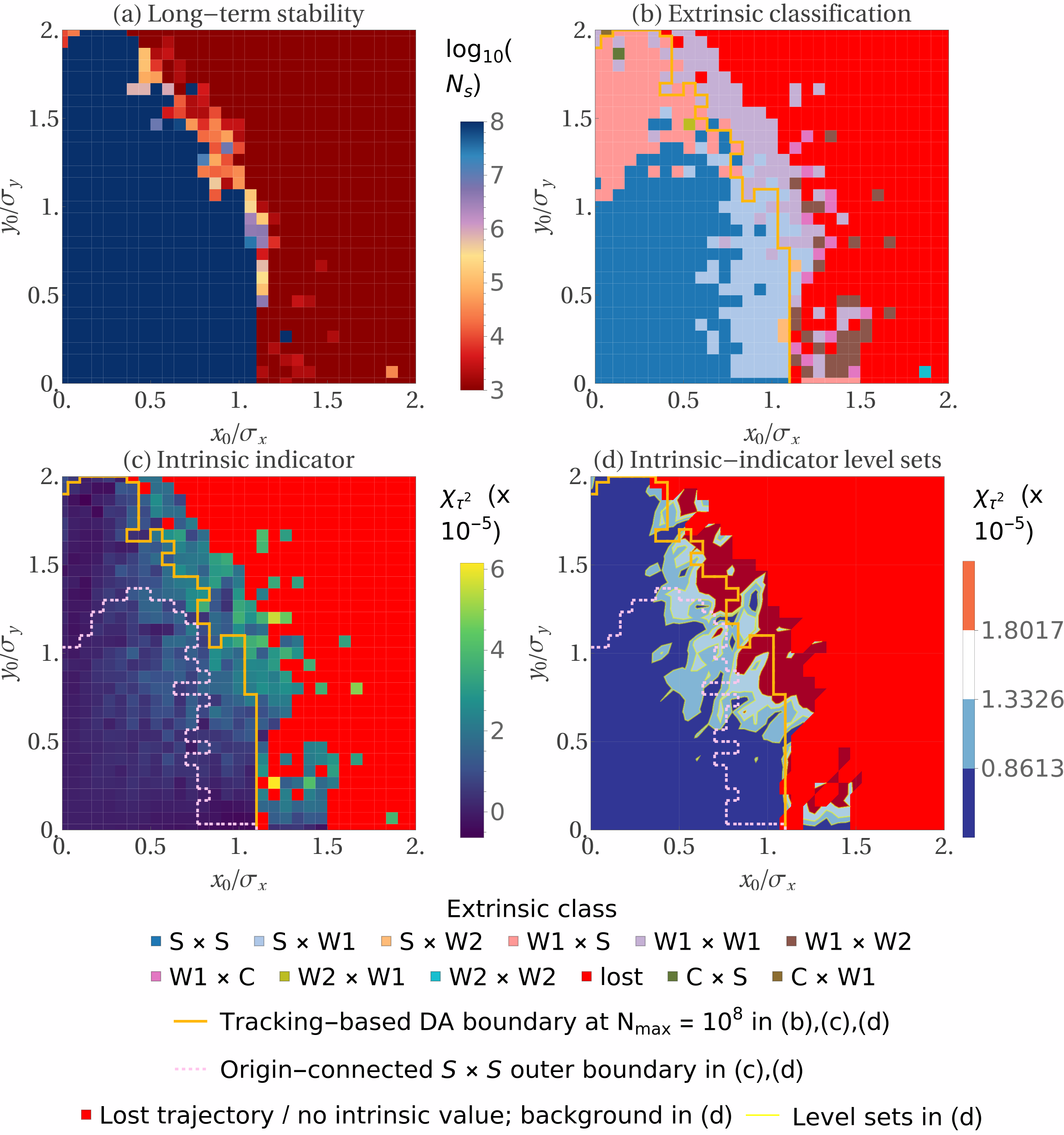}
  \caption{Proton transverse-launch stability diagnostics for the
nonlinear IOTA lattice configuration considered here. (a) Combined long-term stability index
$\log_{10}(N_s)$ using tracking data through $10^8$ turns. (b)  Classification obtained using the nonlinear transformation $\mathcal{A}$ and classifying the two
symplectic-plane projections. (c) Intrinsic
indicator $\chi_{\tau^2}$ computed in linearly scaled coordinates, and (d) its linearly interpolated level sets at the 60th, 70th, and 80th percentiles. All PH diagnostics
are computed from the first 200 turns.}
  \label{fig:TransverseProton}
\end{figure*}

In contrast to the electron case, Fig.~\ref{fig:TransverseProton} (b) shows that the proton configuration has
a more restricted origin-connected $\mathrm{S}\times\mathrm{S}$ region. As in the electron case, for the sampled ICs, this region typically corresponds to smaller values of
$\chi_{\tau^2}$ and lies entirely within the region that survives
through the full $10^8$-turn tracking interval. Around it, the extrinsic classification transitions mainly through $\mathrm{S}\times\mathrm{W1}$ and $\mathrm{W1}\times\mathrm{S}$ toward $\mathrm{W1}\times\mathrm{W1}$, isolated $\mathrm{S}\times\mathrm{S}$ and other mixed labels. Across the tracking-based DA, $\chi_{\tau^2}$ becomes more heterogeneous, particularly where several extrinsic classes meet in panel (b), see panels (c), (d). The surviving portion of these surrounding regions shrinks as the tracking horizon is extended, although several weakly chaotic ICs persist for intermediate durations before loss, consistent with long-lived sticky motion. Thus,
the PH diagnostics computed from only 200 orbit points remain consistent
with the survival structure for a different particle species over
$10^8$ turns, at the upper end of the practical range for computational
long-term proton tracking.

Taken together, the electron and proton results show how the extrinsic classifier complements the intrinsic indicator. For each orbit, the intrinsic indicator assigns a scalar value derived from the PH information of its full 4D point cloud, whereas the extrinsic classifier assigns a categorical, plane-resolved description of the projected dynamics. Although the resulting spatial distributions show broad agreement with each other and with long-term tracking within the sampled region of interest, they are not pointwise equivalent because they summarize distinct topological information. Their combined use can therefore provide a more informative short-orbit characterization for guiding dynamic-aperture studies.

\section{\label{sec:Conclusion and Outlook}Conclusion and Outlook}

We have introduced persistent homology chaos diagnostics for Hamiltonian orbit data, 
with emphasis on beam-physics applications. 
For
2D orbit data, the Euler--Betti classifier distinguishes
invariant tori, island chains, thin chaotic layers, and thick chaos from
short trajectory segments. For 4D phase spaces, we
introduced two complementary constructions: the intrinsic Euler-type
persistence indicators $\chi_{\tau}$ and $\chi_{\tau^2}$, computed from
lifespan sums in homological dimensions zero through two, and an
extrinsic classification obtained by applying the 2D
classifier separately to the two symplectic-plane projections in
normal-form coordinates.

For the canonical examples investigated, these PH diagnostics reproduce the
expected qualitative organization of regular, resonant, weakly chaotic,
and strongly chaotic motion from short orbit segments. Comparisons with
classical chaos indicators support the 2D classification,
while the Hénon--Heiles example demonstrates that the coordinate
representation can affect the recovery of thin loop-like structures.
In
the IOTA applications, the intrinsic and extrinsic diagnostics computed
from the first 200 turns are consistent with the long-term
survival structure for both electron and proton cases. In both configurations, the connected
$\mathrm{S}\times\mathrm{S}$ region containing the reference particle typically
corresponds to lower values of $\chi_{\tau^2}$ and is inside the principal
long-lived region.

From a beam-physics perspective, the IOTA study suggests a staged
workflow. Midplane symmetry permits a flat-beam scan that identifies the
extent of the nested-torus region in the horizontal phase space and
provides a physically motivated region of interest for the transverse
launch. Within this region, the intrinsic indicator and extrinsic PH classifier provide continuous full-orbit and categorical plane-resolved summaries, respectively, which can guide denser sampling toward regions requiring longer tracking. In principle, this
screening stage can avoid uniformly long tracking of every IC by reserving the most expensive calculations for the candidate
boundary and other uncertain regions, although the resulting
computational savings remain to be quantified in a large-scale
DA study. Similar considerations apply to exhaustive DA-optimization campaigns in the design-phase.

The principal limitations arise from the finite representation of each
orbit. The PH results depend on the number of sampled points, the
coordinate representation, the metric, and the persistent-homology
construction. The number of points required is orbit dependent: a
well-sampled, geometrically simple orbit may be resolved using relatively
few points, whereas a more complicated or sparsely sampled orbit may
require more. Across the validation examples, $N_{\mathrm{PH}}$ ranges from 128 to
350 points, while all IOTA
applications use $200$ turns per IC and produce results consistent
with the long-term survival structure. These results show that short
orbit segments can be informative in the two- and four-dimensional
systems considered here, but they do not establish that comparable
sample sizes are sufficient in higher dimensional phase spaces. In six
dimensions and above, more orbit points may be required to resolve the
relevant structures. Extending the present methods to these dimensions
and determining suitable sampling requirements are left for future work.
The intrinsic indicators also require local calibration and do not
provide universal thresholds for chaos. The extrinsic classification
additionally assumes that the motion is uncoupled or can be represented
in approximately decoupled normal-form coordinates, and the projected
labels do not uniquely determine the topology of the full
4D orbit.

Possible directions for future work include improving robustness with
respect to the metric, coordinate representation, and choice of
filtration. This could include testing witness or ellipsoidal
complexes~\cite{CarlssonVJ2021TDA,Sara2024Ellipsoids,Sara2026ellipsoids}, changing the
metric used in the Vietoris--Rips construction~\cite{Helene2026}, or
incorporating directional information from the orbit flow. For example,
flow-aware ellipsoidal filtrations have recently been proposed for the
persistent-homology analysis of recurrent
signals~\cite{Eryilmaz2026}.

Another possible direction is to incorporate short-orbit PH diagnostics
into large-scale six-dimensional DA studies. Such studies involve hundreds of thousands of ICs and may require
tracking for at least $10^6$ turns~\cite{Scandale1995DynamicAperture,ChaosIndicators}. PH computed from short initial orbit segments could provide an early
diagnostic while long-term tracking remains the reference stability
calculation. Its relationship to numerical integrals of motion in
normal-form coordinates could also be
studied~\cite{belanger2025numericalevaluationintegralsmotion}.
There is no unique extension of the present framework to six dimensions. One possibility is a
full-dimensional intrinsic indicator, extended through homological
dimension three; this would include all three degrees of freedom in the
persistence calculation but would remain a scalar diagnostic rather than
a topological classification. When normal-form coordinates approximately
separate the dynamics, alternatives include a 4D
transverse analysis together with a 2D longitudinal
analysis, or three plane-resolved 2D classifications.
Determining which representation is most informative, and whether short
orbit segments remain sufficient, requires further study.

Finally, PH features could be incorporated into hybrid machine-learning
approaches. Recent artificial-intelligence methods have been applied to
the classification of regular and chaotic
motion~\cite{Chaos-ML1,Chaos-DeepLearning}, while persistence information
can be vectorized or represented as persistence images for use as
structured classifier inputs~\cite{Adams2017PersistenceImages}. Similar
PH-to-machine-learning pipelines have been used in condensed-matter
applications, where persistence images were used to identify phase
transitions and construct interpretable order
parameters~\cite{PHinCondensedMatter}. Mapper-based TDA could also be applied to
higher-dimensional symplectic orbit data~\cite{MapperPaper,Lum2013ShapeTopology}.
Whereas the Euler--Betti classifications and Euler-type persistence
indicators introduced here provide quantitative orbit diagnostics,
Mapper could provide a qualitative view of the organization and
connectivity of the orbit data.

\section*{Data Availability}

The data supporting the findings of this study were generated
numerically. The COSY INFINITY data used in the IOTA analysis,
together with additional numerical details for reproducing the
synthetic benchmark data, are available from the corresponding
author upon reasonable request.

\appendix

\section{\label{sec:PH_appendix}Algebraic Topology}

The main text interprets Betti numbers geometrically as counts of
connected components, loops, and voids. These quantities predate their
modern algebraic formulation, in which, over a field, they are the
dimensions of homology vector spaces~\cite{siersma2012poincare,Hatcher}.
This appendix develops that formulation through the classical homology
of simplicial complexes, then introduces the Euler--Poincar\'e
characteristic and applies the K\"unneth theorem to the product spaces
used in the classifiers.

For finite orbit clouds, PH tracks topology across
geometric scales using a Vietoris--Rips filtration. Although the
Vietoris complex dates back roughly a century~\cite{VietorisBiography},
efficient computational implementations of PH are
comparatively recent. Hence, we introduce the barcodes, Euler--Betti
curves, lifespan indicators, and stability bound used in the main text.

\subsection{Motivation and classical homology}

One of the earliest and most successful classification ideas in
algebraic topology is already present in Euler's classical formula for
polyhedra,
\begin{equation}
\chi_{\mathrm{classical}}(X)=V-E+F,
\label{eq:euler_classical}
\end{equation}
where $X$ is a polyhedral surface and $V$, $E$, and $F$ denote its
numbers of vertices, edges, and faces, respectively. The key point is not the particular shape of the polyhedron, but
that the combination $V-E+F$ remains unchanged under continuous
deformations that preserve its topology. This provides an early example
of a topological invariant: a possibly complicated space is assigned
algebraic information that does not depend on its particular
representation.

Classical homology generalizes this idea by assigning an abelian group
to a topological space in each nonnegative dimension. Several standard
constructions yield equivalent homological information~\cite{Hatcher}. Because the
Vietoris--Rips complexes used in this work are simplicial complexes, we
introduce the required ideas through simplicial homology.

An $n$-simplex has $n+1$ vertices. Thus, a
$0$-simplex is a vertex, a $1$-simplex is an edge, a $2$-simplex is a
filled triangle, and a $3$-simplex is a filled tetrahedron. A simplicial
complex is a collection of simplices that contains all faces of each
simplex and in which two simplices, when they intersect, do so along a
common face. 

To combine simplices algebraically, we choose a coefficient field
$\mathbb{F}$. For a simplicial complex $X$, the $n$-th chain group
$C_n(X;\mathbb{F})$ is the $\mathbb{F}$-vector space consisting of
finite linear combinations of oriented $n$-simplices,
\begin{equation}
c=\sum_i a_i\sigma_i,
\qquad a_i\in\mathbb{F}.
\label{eq:n_chain}
\end{equation}

The linear boundary operator
\begin{equation}
\partial_n:C_n(X;\mathbb{F})
\longrightarrow C_{n-1}(X;\mathbb{F})
\label{eq:boundary_operator}
\end{equation}
maps an $n$-dimensional chain to its $(n-1)$-dimensional boundary. For
an oriented simplex, it is defined by
\begin{equation}
\partial_n[v_0,\ldots,v_n]
 =
\sum_{i=0}^{n}(-1)^i
[v_0,\ldots,\widehat{v_i},\ldots,v_n],
\label{eq:simplex_boundary}
\end{equation}
where $\widehat{v_i}$ means that the vertex $v_i$ is omitted.  For
example, $\partial_1[v_0,v_1]=[v_1]-[v_0]$, so the boundary consists of
the two endpoints, with signs determined by the chosen orientation For computational purposes, PH
is often calculated using coefficients in
$\mathbb{F}=\mathbb{Z}_2=\{0,1\}$.
In this field, $-1\equiv1$, so subtraction and addition are
identical and the choice of orientation does not affect the computation.

Successive boundary operators satisfy
\begin{equation}
\partial_{n-1}\circ\partial_n=0,
\label{eq:boundary_squared_zero}
\end{equation}
which expresses the familiar statement that the boundary of a boundary
vanishes. This property leads naturally to two important groups. An
$n$-chain with no boundary is called an $n$-cycle, and the group of all
$n$-cycles is
\begin{equation}
Z_n(X;\mathbb{F})=\ker\partial_n.
\label{eq:cycle_group}
\end{equation}
An $n$-chain that is itself the boundary of an $(n+1)$-chain is called
an $n$-boundary, and the group of all $n$-boundaries is
\begin{equation}
B_n(X;\mathbb{F})=\operatorname{im}\partial_{n+1}.
\label{eq:boundary_group}
\end{equation}
Equation~\eqref{eq:boundary_squared_zero} implies that every boundary is
also a cycle,
$B_n(X;\mathbb{F})\subseteq Z_n(X;\mathbb{F})\subseteq C_n(X;\mathbb{F})$. Not every cycle is necessarily a boundary. A closed loop surrounding a
hole, for example, is a cycle but cannot be the boundary of a
two-dimensional chain contained in the space. Homology records precisely
this difference. The $n$-th homology group is the quotient
\begin{equation}
H_n(X;\mathbb{F})
 =
Z_n(X;\mathbb{F})/B_n(X;\mathbb{F}).
\label{eq:homology_group}
\end{equation}

This quotient is a set of equivalence
classes. A class is trivial when its representative cycle
is itself a boundary. A nontrivial class represents a cycle that cannot
be filled by a higher-dimensional chain contained in the space.
Informally, these nontrivial classes describe the independent
$n$-dimensional holes of $X$.

\subsection{\label{sec:Betti,Euler-Poincare,Kunneth}Betti numbers, the Euler--Poincaré characteristic, and the K\"unneth theorem}

Throughout this work, homology is computed with coefficients in
$\mathbb{F}=\mathbb{Z}_2$. With this choice, the chain, cycle, boundary,
and homology groups can be viewed as vector spaces over
$\mathbb{Z}_2$. To simplify the notation, the coefficient field will
be suppressed in what follows.

The \emph{Betti numbers} are defined as
\begin{align}
\beta_n(X)&\equiv\dim H_n(X)\\
&=\dim Z_n(X)-\dim B_n(X).
\end{align}
The Betti numbers used in this work are interpreted as $\beta_0(X)$, the number of
connected components of $X$; $\beta_1(X)$, the number of independent
loops of $X$; and $\beta_2(X)$, the number of independent voids of $X$.

For a finite simplicial complex, the classical Euler characteristic is
defined by the alternating sum of the chain-group dimensions. The
Euler--Poincar\'e formula~\cite{Hatcher,EdelsbrunnerHarer2010CompTopo, CarlssonVJ2021TDA} states that it can equivalently be computed
from homology. In this work, we use its Betti-number form and write

\begin{align}
\chi(X)\equiv\chi_{\mathrm{classical}}(X)
&=\sum_{n=0}^{d}(-1)^n\dim C_n(X)\\
&=\sum_{n=0}^{d}(-1)^n\dim H_n(X)\\
&=\sum_{n=0}^{d}(-1)^n\beta_n(X).\label{Euler-Poincare}
\end{align}

The preceding quantities can also be used to study spaces constructed
from simpler spaces. Given topological spaces $X$ and $Y$, we can construct a new topological space $X\times Y$ given by the Cartesian product of $X$ and $Y$ equipped with the product topology. The K\"unneth theorem describes how
the homology of the resulting product space $X\times Y$ is related to
the homology of its factors $X$ and $Y$. Since homology is computed
over the field $\mathbb{Z}_2$ in this work, the K\"unneth theorem takes
the form
\begin{equation}
H_n(X\times Y)
\cong
\bigoplus_{i+j=n} H_i(X)\otimes H_j(Y),
\label{Kunneth Product}
\end{equation}
where $\oplus$ denotes a direct sum and $\otimes$ denotes the tensor
product~\cite{Hatcher}. Taking dimensions on both sides gives
\begin{align}
\beta_n(X\times Y)
&=
\dim H_n(X\times Y)\\
&=
\sum_{i+j=n}
\dim\!\left(H_i(X)\otimes H_j(Y)\right)\\
&=
\sum_{i+j=n}\beta_i(X)\beta_j(Y).
\label{Betti Product}
\end{align}
Thus, the Betti numbers of a product space can be computed from the
Betti numbers of its factors. Consequently,
\begin{align}
\chi(X\times Y)
&=\sum_n(-1)^n\sum_{i+j=n}\beta_i(X)\beta_j(Y)\\
&=\sum_{i,j}(-1)^{i+j}\beta_i(X)\beta_j(Y)
=\chi(X)\chi(Y).
\label{Euler-Poincare Product}
\end{align}

These classical-homology results provide the idealized topological basis
for the two- and four-dimensional classifiers developed in the main
text. Classical homology considers a topological space as a whole, such
as an entire circle or two-torus. For the short numerical orbits
considered in this work, however, only a finite subset of such an
idealized space is accessible. The following subsection therefore
introduces persistent homology, which provides the computational
framework used to extract topological information from finite orbit samples.

\subsection{\label{sec:PH_persistent}Persistent homology}

To make homology computationally applicable to finite data, it is useful to view the construction as an evolution of topological spaces. The initial space $X_0$ is the point cloud associated with the original data. If it contains $N$ distinct points, then each point is an isolated connected component, so that
\begin{equation}
\beta_0(X_0)=N,
\quad
\beta_n(X_0)=0
\quad \text{for } n>0.
\end{equation}

The initial space $X_0$ is then evolved with respect to a filtration parameter (geometric scale) $\epsilon\geq0$, producing the nested sequence
\begin{equation}
X_0
\subseteq
X_{\epsilon_1}
\subseteq
\cdots
\subseteq
X_{\epsilon_{\max}},
\quad
0<\epsilon_1<\cdots<\epsilon_{\max}.
\end{equation}
The complete evolution is denoted by $\{X_\epsilon\}$ and the corresponding homology groups by $H_n(X_\epsilon)$. In this analogy, $\epsilon$ plays the role of an evolution parameter and the Betti numbers $\beta_n(X_\epsilon)$ are observables that describe the changing topology of the system. 

There are several ways to generate the filtration $\{X_\epsilon\}$ that
defines this topological evolution. In this work, we use the
Vietoris--Rips construction, computed with the
Python package \texttt{ripser.py}~\cite{Tralie2018}. Given any metric $d$ on the finite point
cloud $X_0$, the Vietoris--Rips complex at scale $\epsilon$ is defined by
\begin{equation}
\begin{aligned}
X_\epsilon\equiv\operatorname{VR}_\epsilon(X_0)=
\left\{
[v_0,\ldots,v_n]: v_i\in X_0,
d(v_i,v_j)\leq\epsilon,\ \forall\; 0\leq i<j\leq n
\right\}.
\end{aligned}
\end{equation}
Thus, an edge is added when the distance between its two vertices is at
most $\epsilon$, and a filled triangular face is added when this condition
holds for all three pairs of its vertices. More generally, an $n$-simplex
is added whenever every pair of its $n+1$ vertices lies within distance
$\epsilon$. When $d$ is the Euclidean metric, this can be visualized by
placing closed balls of radius $\epsilon/2$ around the points: an edge is
created when two balls intersect, and a higher-dimensional simplex is
created when all corresponding pairs of balls intersect. The topology of $X_\epsilon$ therefore evolves with the filtration: connected components merge, loops are created and filled, and higher-dimensional cavities may appear and disappear.

Persistent homology records these topological transitions throughout the evolution $\{X_\epsilon\}$. The $k$th homology class in $H_n(X_\epsilon)$ is born at the filtration value $b_{n,k}$ at which it first appears and dies at the filtration value $d_{n,k}$ at which it ceases to persist. This information is represented by the $n$th barcode
\begin{equation}
B_n=\left\{(b_{n,1},d_{n,1}),
\ldots,(b_{n,N_n},d_{n,N_n})\right\}.
\end{equation}
The collection of barcodes through homological dimension $d$ is denoted by
$\mathbf{B}=\{B_0,\ldots,B_d\}$. The same information may also be represented by the corresponding PD, in which each class is represented by the point $(b_{n,k},d_{n,k})$.

Examples of PDs appear in the (b) panels of Figs.~\ref{fig:Torus}--\ref{fig:ThickChaos}. In Fig.~\ref{fig:Torus}, the PD of the torus class contains blue points from $B_0$, and a single orange point from $B_1$, reflecting
one dominant loop in $H_1$. Long bars, or points far from the diagonal $b=d$,
represent topological structure that persists over a wide range of
filtration scales. Short bars, or points close to the diagonal, represent features that persist over only a small range of filtration scales. Such features are sometimes associated with noisy data but do not necessarily represent noise. For example, the short orange bars in the PD of 
Fig.~\ref{fig:Island} correspond to the genuine smaller-scale loops of the island chain, whereas the additional short orange bars in the PD of Fig.~\ref{fig:ThinChaosT1} reflect the loop structure associated with thin chaos.

To study Hamiltonian orbits generated by a symplectic map, we use the following Euler-characteristic-type quantities:
\begin{align}
E(\epsilon)\equiv\chi(X_\epsilon) &=\sum_{n=0}^{d} (-1)^n \beta_n(X_\epsilon),\label{Euler-Betti}\\
\chi_{\tau^p}(\mathbf{B})
&=
\sum_{n=0}^{d}(-1)^n
\tau^p(B_n),
\label{eq:chi_taup}
\end{align}
where for $p>0$, we define the $p$-th lifespan sum in homological dimension $n$ as
\begin{equation}
\tau^p\left(B_n\right)=\sum_k \ell_{n,k}^p=\sum_k \left(d_{n,k}-b_{n,k}\right)^p,
\label{eq:tau_np}
\end{equation}
where the sum is taken over all finite PH classes in dimension $n$. In the main text, we write $\beta_n(\epsilon)$ for the Betti number $\beta_n(X_\epsilon)$ appearing in Eq.~\eqref{Euler-Betti}. For the examples in Figs.~\ref{fig:Torus}--\ref{fig:ThickChaos}, the
individual Betti curves $\beta_0(\epsilon)$ and $\beta_1(\epsilon)$ are
shown in the (c) and (d) panels, while
the corresponding Euler--Betti curve $E(\epsilon)$ in (c).

The Euler--Betti curve $E(\epsilon)$ records the alternating sum of the Betti numbers at each filtration value, but it does not directly account for how long each individual homology class persists. In contrast, $\chi_{\tau^p}$ incorporates the lifespan of each finite homology class. We use $p=1$ and $p=2$ as our main lifespan summaries, with $p=2$ assigning greater relative weight to longer-lived classes. These Euler-type quantities are analogues of the Euler--Poincar\'e characteristic, the case $p=1$ first introduced in Ref.~\cite{Adler_2010}.

Finally, we establish the stability bound stated in
Sec.~\ref{sec:Intrinsic}. Let $P$ and $Q$ be finite point clouds, and let
$W_p$ denote the $p$-Wasserstein distance between
PDs~\cite{CarlssonVJ2021TDA}, using the $\ell^\infty$ ground metric and
allowing points to be matched to the diagonal. For matched diagram
points $x=(b,d)$ and $y=(b',d')$, including diagonal points for which
$\ell=0$, the lifespan function $\ell(b,d)=d-b$ satisfies
$|\ell(x)-\ell(y)|\leq2\lVert x-y\rVert_\infty$. The reverse triangle
inequality therefore bounds the difference of the corresponding
$p$th-root lifespan sums by
$2W_p(B_n(P),B_n(Q))$. For $p>1$, applying
$|a^p-b^p|\leq p\max\{a,b\}^{p-1}|a-b|$ and summing over the
homological dimensions yields

\begin{equation}
\begin{aligned}
\left|\chi_{\tau^p}(P)-\chi_{\tau^p}(Q)\right|
&\leq
2p\sum_{n=0}^{d}
\max\!\left\{
\tau^p(B_n(P)),\tau^p(B_n(Q))
\right\}^{(p-1)/p}\\
&\qquad\quad\times
W_p\!\left(B_n(P),B_n(Q)\right).
\end{aligned}
\end{equation}

For $p=1$, the same argument gives
\begin{equation}
\left|\chi_{\tau}(P)-\chi_{\tau}(Q)\right|
\leq
2\sum_{n=0}^{d}
W_1\!\left(B_n(P),B_n(Q)\right).
\end{equation}

\bibliography{Ph}

\end{document}